%% file: main.tex
\documentclass{article}

\usepackage{PRIMEarxiv}

\usepackage[utf8]{inputenc} 
\usepackage[T1]{fontenc}    
\usepackage[colorlinks=true,linkcolor=blue,anchorcolor=blue,citecolor=blue,filecolor=blue,menucolor=blue,runcolor=black,urlcolor=black]{hyperref}       
\usepackage{url}            
\usepackage{booktabs}       
\usepackage{amsfonts}       
\usepackage{nicefrac}       
\usepackage{microtype}      
\usepackage{lipsum}
\usepackage{fancyhdr}       
\usepackage{graphicx}       
\usepackage{authblk}

\usepackage{multirow}

\usepackage{subcaption}
\usepackage{float}
\usepackage{amsmath}
\usepackage{algorithm}
\usepackage{algpseudocode}
\usepackage{amssymb}
\usepackage{mathtools}
\usepackage{tabularx}
\usepackage{colortbl}
\usepackage{eurosym}

\usepackage{comment}
\usepackage{csquotes}

\title{Improved Performances and Motivation\\ in Intelligent Tutoring Systems:\\ Combining Machine Learning and Learner Choice

}

\author[1,$\times$,*]{Benjamin Clément}
\author[1,2,+]{Hélène Sauzéon}
\author[1]{Didier Roy}
\author[1,+,$\times$]{Pierre-Yves Oudeyer}
\affil[1]{Inria, FLOWERS team, Talence, 33405, France}
\affil[2]{Université de Bordeaux, BPH lab, Bordeaux, 33076, France}
\affil[+]{these authors equally supervised this work}
\affil[*]{work presented here was done while at Inria}
\affil[$\times$]{corresponding authors: benji.clement@gmail.com, pierre-yves.oudeyer@inria.fr}

\begin{document}
\maketitle

\begin{abstract}


Large class sizes challenge personalized learning in schools, prompting the use of educational technologies such as intelligent tutoring systems. 
To address this, we present an AI-driven personalization system, called ZPDES, based on the Learning Progress Hypothesis - modeling curiosity-driven learning - and multi-armed bandit techniques. It sequences exercises that maximize learning progress for each student. 
While previous studies demonstrated its efficacy in enhancing learning compared to hand-made curricula, its impact on student motivation remained unexplored. Furthermore, ZPDES previously lacked features allowing student choice, a limitation in agency that conflicts with its foundation on models of curiosity-driven learning.

This study investigates how integrating choice, as a gamification element unrelated to exercise difficulty, affects both learning outcomes and motivation. We conducted an extensive field study (265 7-8 years old children, RCT design), comparing ZPDES with and without choice against a hand-designed curriculum. Results show that ZPDES improves both learning performance and the learning experience. Moreover adding choice to ZPDES enhances intrinsic motivation and further strengthens its learning benefits. In contrast, incorporating choice into a fixed, linear curriculum negatively impacts learning outcomes.

These findings highlight that the intrinsic motivation elicited by choice (gamification) is beneficial only when paired with an adaptive personalized learning system. This insight is critical as gamified features become increasingly prevalent in educational technologies.

\end{abstract}


\clearpage

\input{intro}

\input{results}

\input{discuss}

\input{methods}


\section{Acknowledgements}

We thank the teachers of the Bordeaux School District for providing access to their classrooms, an essential component for our empirical study, as well as the Rectorate of Bordeaux-Nouvelle Aquitaine, a regional administrative division of the French Ministry of Education, and its Digital Education mission, with whom Inria established a partnership convention, enabling our access to educational institutions.
Special acknowledgment is given to Josias Levi Alvarez and Medhi Alaimi, two internship students, who were pivotal in conducting the in-class experiments, significantly contributing to the success of our study. This project benefited from funding from ANR AI Chair DeepCuriosity ANR-19-CHIA-0004.

\input{Appendix}
\bibliographystyle{unsrt}  
\bibliography{references}

\end{document}

%% file: intro.tex

\section{Introduction}

A key challenge of 21st century schools is to make students active and engaged in their education with the difficulty of dealing with a wide diversity of students' abilities and motivations for learning. The growing research on personalized or individualized education as well as on active teaching testifies to this huge societal need targeting the equality of opportunities at school for all\cite{bartolome2018personalisation}.

The evidence-based assets of personalized learning over one-size-fits-all educational approaches are today well documented \cite{deunk2015differentiation,  iterbeke2021effects}. As classroom sizes are still high, it is difficult for teachers to set up individualized teaching paths, which is why high expectations are placed on Educational Technologies (ET) to automate them and support teachers in their missions.
The exploitation of artificial intelligence and digital systems has then become a crucial question to improve ET \cite{collins2018rethinking} and enable forms of adaptivity and personalization, leading to the development of Intelligent Tutoring Systems (ITS). This aims to make education more effective and accessible for the large diversity of students and as a way to provide useful objective metrics on learning \cite{anderson1995cognitive, koedinger1997intelligent, nkambou2010advances}. 

Among the ITS field, there has been several approaches to optimize, personalize and adapt ITS to learners in order to enhance access to quality learning experience for all learners. 
Adaptation in learning technologies can be described as a structure with three main components \cite{vandewaetere2011contribution}. 
Firstly, adaptation to the instruction source, i.e. to what it will be adapted, such as the learner learning style \cite{sun2007use, bunderson2000building}, knowledge \cite{koedinger1993effective} or preferences \cite{ray2007teaching}. Secondly, the target of the adaptive instruction, i.e. what will be adapted, such as the content \cite{sun2007use} or the presentation \cite{milne1997adapting}. Thirdly, the adaptive component generating a pathway between the two first components, i.e. how to adapt a Target to a Source, such as rule-based systems \cite{sun2007use} or Bayesian-networks \cite{conati2002using}. 
This last component is the engine generating a curriculum of training activities for learners in ITS. Thus, to be able to adapt the content to the learner, all ITS are most often implicitly or explicitly based on the concepts of the zone of proximal development (ZPD) \cite{vygotsky1978mind} and the state of Flow \cite{csikszentmihalyi1975beyond}. 

These are well-known concepts in developmental psychology, and have inspired many models of learning for education (e.g., Cognitive load theory\cite{sweller2011cognitive}).  
Following them, many ITS aim to offer the learner pedagogical activities that are neither too difficult nor too easy with regard to their abilities, so that they can be engaged and progress in their acquisitions without being anxious or bored during the process. ITS can also propose activities the learner can not solve alone but will be able to solve with hints or with the teacher's help. 
From it, the ITS can be divided in two main design categories for managing learning curricula \cite{bartolome2018personalisation}. The former, namely the "linear design" involves that all learners follow a single path although at different speeds or with a different number of attempts. The latter, called "branched-paths design", enables each learner to follow a specific path according to their their own need.

Recently, with the growing interest in the phenomenon of curiosity\cite{murayama2019reward}, these concepts have been revisited in the Learning Progress (LP) model \cite{kaplan2007search,oudeyer2016intrinsic}. In this contemporary reward-learning model,  curiosity-based intrinsic motivation and learning progress are linked by a virtuous loop: 
the child learns better on tasks for which she is interested and intrinsically motivated and, in return, the learning progress yielded generates an internal reward that stimulates his intrinsic motivation to continue acquiring knowledge for its own sake \cite{gottlieb2013information}; (see also Self-determination theory\cite{ryan2020intrinsic}). In other words, this model stresses the personal factors in learning, where individual LP contributes to both learner's motivation and  self-organisation of its active exploration of learning tasks\cite{ten2021humans}. Furthermore, machine learning research has shown that using LP to automatically generate learning curricula, by sampling tasks with maximal expected LP, is a powerful heuristic that leads to sample efficient learning of skills and world models \cite{oudeyer2007intrinsic2, schmidhuber1991curious, lopes2012exploration, graves2017automated, colas2019curious, kim2020active, portelas2020automatic}. In other words, when one aims to maximize long-term learning outcomes on a variety of tasks, LP can be used as an efficient proximal heuristic for sampling learning activities. Thus, the LP model argues that LP-based learning curricula shall lead both to intrinsically motivating and efficient long-term learning. 

\begin{figure*}[ht!]
\centering
     \centering
            \includegraphics[width=\textwidth]{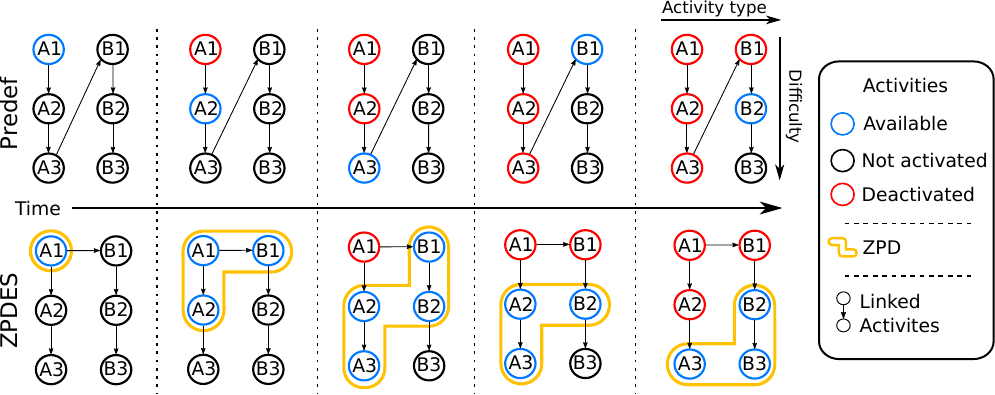}
       \caption{The space of available activities always contain only one activity in the predefined linear sequence (Predef) while the space expends over time with ZPDES to allow a diversity of exploration and find the best activities for the learner. The type of activity is represented by a letter (A or B), and the difficulty by a number (1,2 or 3).}
        
        \label{fig:ZPDESexplore}
\end{figure*}

In this context, the ZPDES algorithm (Zone of Proximal Development and Empirical Success) has been proposed to be used as a new LP-based activity manager in ITS\cite{Clement2014edm}.
Leveraging multi-armed bandit algorithms\cite{auer2003nonstochastic,bubeck2012MAB}, it integrates an expert knowledge rule-based system and exploits the LP to select the activities which present the highest learning value for the student progress.
Basically, ZPDES algorithm uses the student success rate to dynamically computes the best activity set (ZPD) and ponders the activities from LP to select them stochastically. The general idea is illustrated in figure~\ref{fig:ZPDESexplore} and shows the evolution of the possible activities available to the student as  compared to what happens with a predefined linear sequence (Predef). ZPDES has been evaluated as an efficient algorithm for adaptive generation of various learning paths for real students\cite{clement2015}, robust to heterogeneous populations as shown by complementary systematic experiments with simulated learners\cite{clement2016comparison}. 

In summary, the automated personalization performed by ZPDES aims to lead learners into activities providing maximal learning progress, resulting in individualized learning paths. Yet, this approach had so far two major limitations. First, while its associated objective was to enhance intrinsic motivation, the motivational impact of ZPDES was not studies so far on human learners. Second, ZPDES does not lead the learner to actively make decisions related to the learning path although the theoretical LP model\cite{kaplan2007search} encompasses self-decision making. 

Indeed, allowing self-decisions can boost the sense of agency and have a positive motivational impact, and be an efficient vector of performance \cite{leotti2011inherent, murayama2013self, cordova1996intrinsic, DARVISHI2024104967}. One original argument for this design choice for ZPDES relied on findings in educational psychology, in particular those related to the self-regulation model of decision making\cite{byrnes2013nature} revealing that decision making can be biased and error prone, particularly in children\cite{miller2001achieve} which were a priority target of this approach. A child can lack a given resource required to make an adaptive decision (e.g., lacking adequate knowledge) or some factors could constrain the person's ability to carry out decision-making processes (e.g., under-/over-estimation of learner regarding his learning progress)\cite{byrnes2013nature,miller2001achieve}.

\begin{figure}[ht!]
\centering
 
    \begin{subfigure}{0.47\textwidth}

      \includegraphics[width=\textwidth]{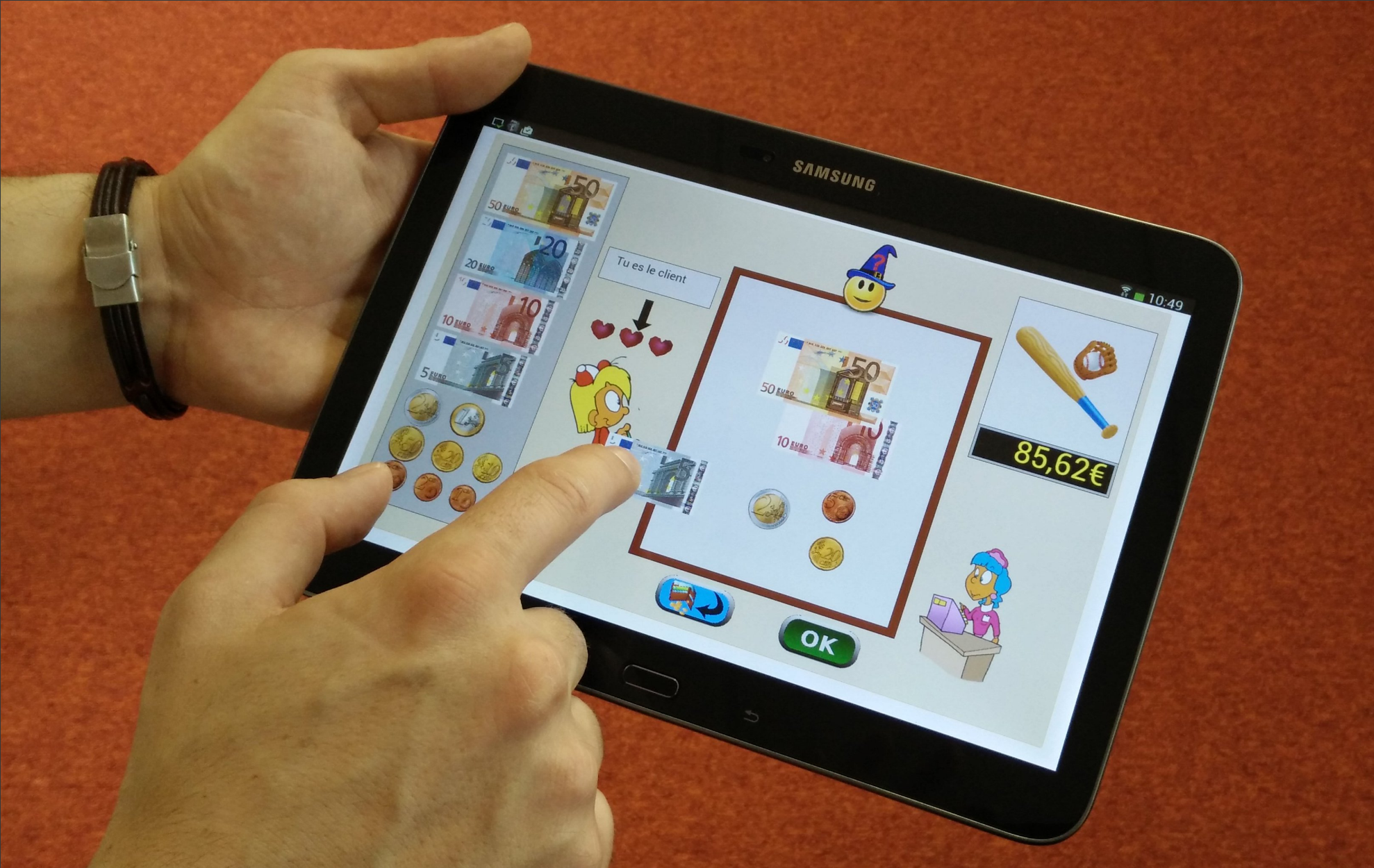}
        \caption{Kidlearn software on tablet}
        
        \label{fig:kidlearnPhoto}
    \end{subfigure}
    \hspace{0.05\textwidth}
    \begin{subfigure}{0.47\textwidth}
            
            \includegraphics[width=\textwidth]{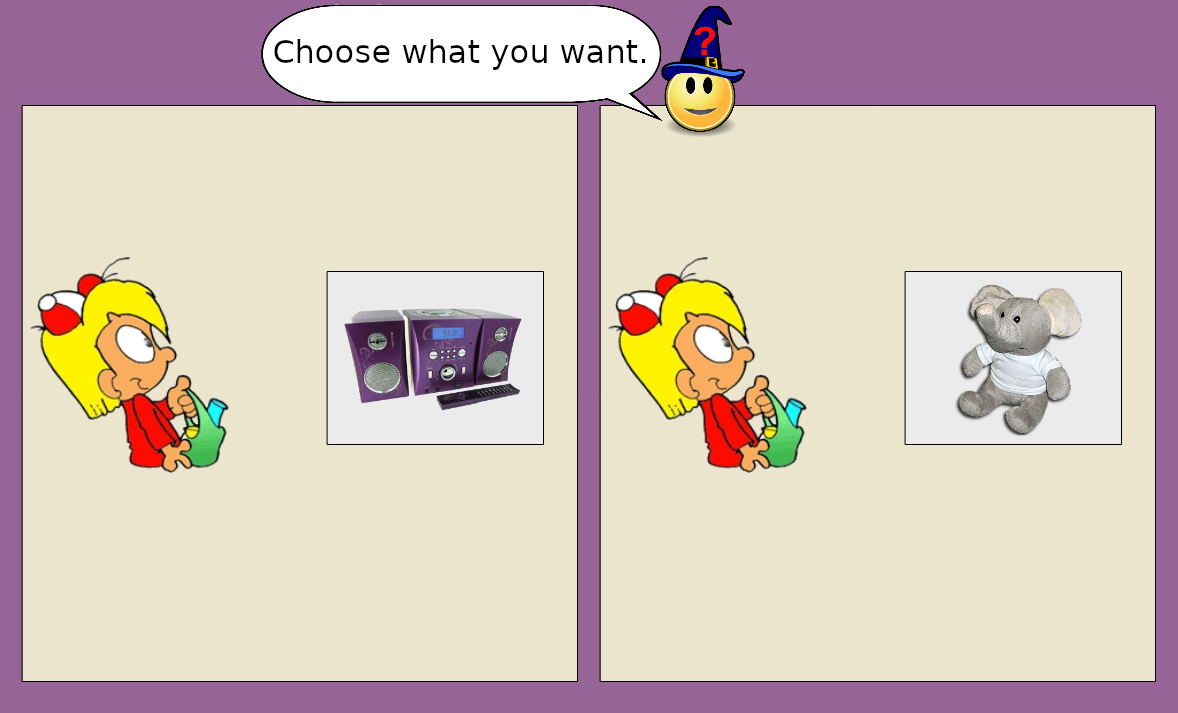} 
        \caption{Contextual Choice given to the student}
        
        \label{fig:kidlearnChoice}
    \end{subfigure}

    \caption[ZpdesVsPredef]{Kidlearn software user interface}

    \label{fig:interface}
\end{figure}

A simple way to overcome potential decision making failures, while promoting the learner's intrinsic motivation to perform the activity, is to combine the use of ZPDES (to control the curriculum difficulty and variety) with offering choice possibilities limited to dimensions that are orthogonal to the learning complexity (e.g. here, choice of visual objects on which to do a math exercise as shown in Fig.~\ref{fig:kidlearnChoice}). This is the principle of ZCO\footnote{ZCO stands for Zone of proximal development with ChOice} system introduced in this paper. Indeed, the subjective value of a task influences academic performance\cite{wigfield1992development, brophy1999toward} and allowing young students (3rd through 9th-grade students\cite{harter1981new}) to express their interest or preferences stimulates their intrinsic motivation and thus could be a booster for ITS effectiveness. An open question we address here is to understand the relative contributions and interactions of LP-based curriculum personalization and choice over both learning efficiency and motivation. 

From the overall data, our first contribution is to show that ZPDES allows human learners to be more motivated and to learn better than a linear design based activity manager (which was made in collaboration with a pedagogical expert in maths teaching) named "Predefined sequence" (Predef), confirming the theoretical LP model and the relationships between learning progress and learner's intrinsic motivations. 
The second contribution is to show the synergistic effect between ZPDES and the ability given to children to express some choices (i.e. to make some decisions which is also a general form of gamification) on both learning performance and motivation. On the contrary, we show that giving children the ability to express some choices in the predefined sequence approach lowers the learning performance. Thus, the effect of choice on learning performance is contextual, and is here positive only for personalized learning curricula.

These two contribution are the result of a field study conducted according to an randomized control trial (RCT) design. Indeed, several systematic reviews reported promising or even positive results on the value-added of ITS\cite{aleven2016instruction,faber2017effects,ma2014intelligent}, while pinpointing methodological limitations of this new empirical field (no control group, no initial group equivalence, no pre- and post-intervention measurements, etc., \cite{CHEUNG201388}) and the great variability of the ITS designs making it difficult to identify which of the ITS features are critical for successful personalized learning\cite{hew2013use, gerard2015automated, bartolome2018personalisation}. 
    
This RCT, approved by Inria COERLE ethical committee, involved 265 children, from 24 classes of 11 primary schools of the Bordeaux school district. The software on which children studied during this RCT, named  Kidlearn ITS\cite{clement2015}, was designed and developed specifically for this kind of experiment. This ITS aims to teach basic mathematics for children aged 7 years old through manipulation of money bills and coins (number decomposition, addition, subtraction of integers and decimals), and has been aligned to official pedagogical objectives on this topic in the national French education system.
Figure~\ref{fig:kidlearnPhoto} shows Kidlearn interface on a tablet the students use. They either play the role of the client or the merchant and need to compose the correct amount of money to either pay or give the change by dragging and dropping the bills and coins on the left side. 
You can refer to \ref{sec:kidlearn} for more details about the pedagogical scenario and interface description.
We compared 4 versions of the KidLearn ITS\cite{clement2015}; the Predefined sequence without (Predef) or with (PCO) learner decision, and ZPDES condition without (ZPDES) or with (ZCO) learner decision.  

%% file: results.tex

\section{Results}
\label{sec:results}

The data presented here involves 265 children (Predef: 62, PCO: 59, ZPDES: 76, ZCO: 68) from schools of the Bordeaux school district. The experiment consisted of four sessions per class over two successive week (two sessions per week with at least one day break between each sessions). During the sessions, students interact with a tablet and either answer questionnaires (pre/post test, motivation questionnaires, etc) or work on Kidlearn ITS for 30 minutes (without planned interruption). The detailed experimental schedule and overall setup are presented in \ref{sec:expProtocol}. 

To study the impact of LP-based personalizing and self-decision-making on student's learning performance and motivation, we use ZPDES as the LP-based algorithm (see \ref{sec:ZPDES}) and a predefined sequence following a "linear design" (called "Predef" described in \ref{sec:PredefSeqSec}) as a baseline. This predefined sequence is implemented as a series of activities in which the student must have 75\% success over 4 activities of the same type to pass to the next activity type. The impact of self-decision making, an assumed in the LP-model, is also studied. It takes the form of a contextual choice given to the student  consisting in choosing the visual objects presented during the exercise, thus the choice does not impact the difficulty of the exercise (details in \ref{sec:choice}). This leads to the comparison of four experimental  conditions, two conditions without self-decision-making: ZPDES and Predef; two conditions with it: ZCO (ZPDES with Choice of Object) and PCO (Predef with Choice of Object).

We first check the differences in curricula, i.e the student progression in the Kidlearn app scenario, between each conditions. To grasp these differences, we compare the learning activities they reach and achieve during training as well as activity space evolution through time. 
Differences in curricula raise the question of the learning effectiveness of each conditions 
and their impact on learning experience and motivation. 
The learning effectiveness is evaluated through comparison between pre- and post-test results, while an emotional scale is used to assess the emotional valence of learning experience and the motivation is evaluated through Vallerand’s questionnaire\cite{vallerand1992academic, vallerand1989construction} (see \ref{sec:measureTool} for more information on measures). Then, the relations between LP-based personalization and subsequent learning performance and motivation are analysed.
Finally, we check if individual characteristics modulate the impact of LP-based personalization.
 
\subsection{How do LP-based personalized curricula differ from hand-designed ones?}
\label{sec:perfResult}
\begin{figure*}[ht]
    \centering
    \hspace*{-0.8cm}
        \begin{tabular}{cc}
            Score for reached activities & Score for achieved activities \\
            \includegraphics[width=0.5\textwidth]{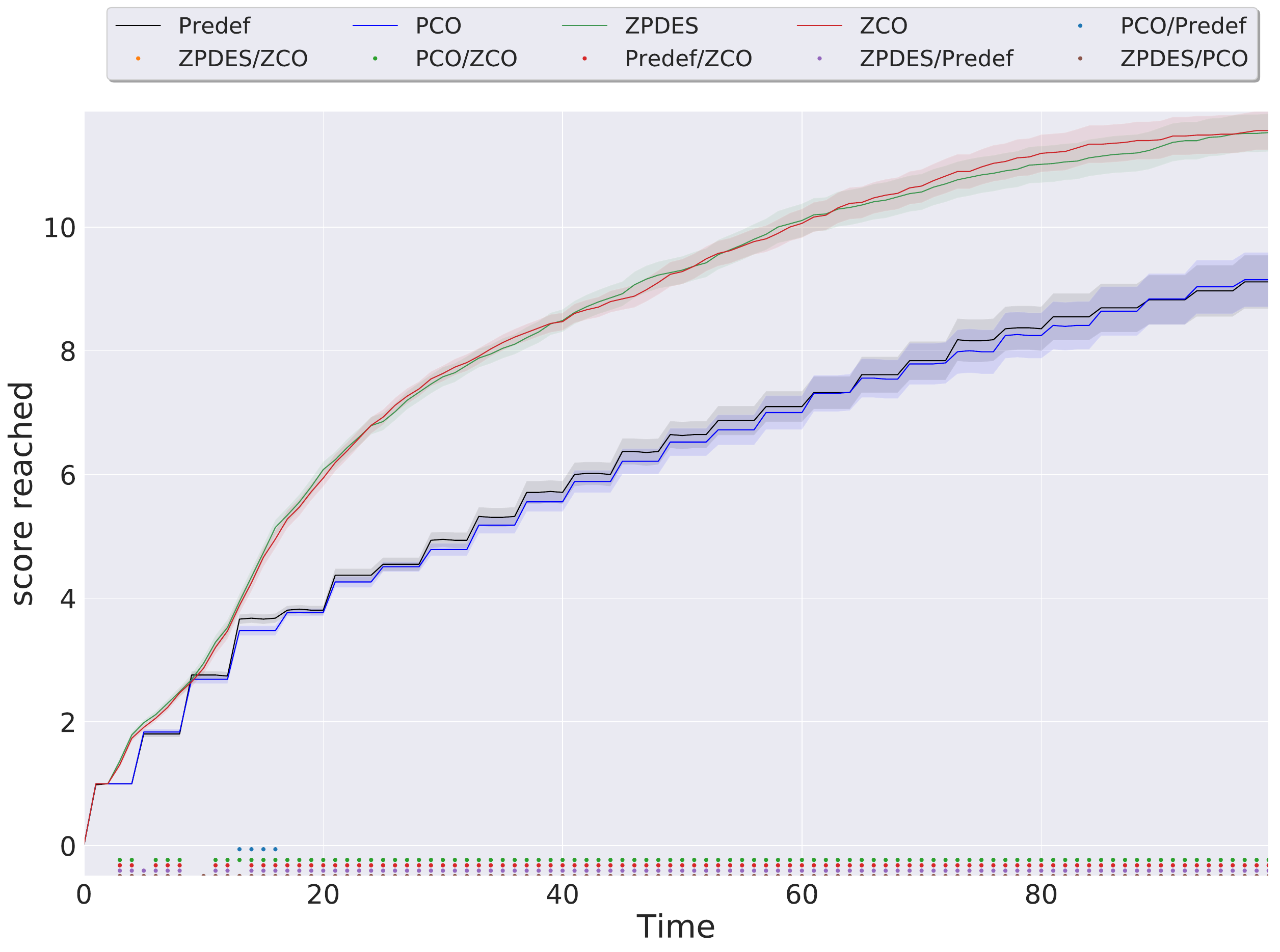} & 
            \includegraphics[width=0.5\textwidth]{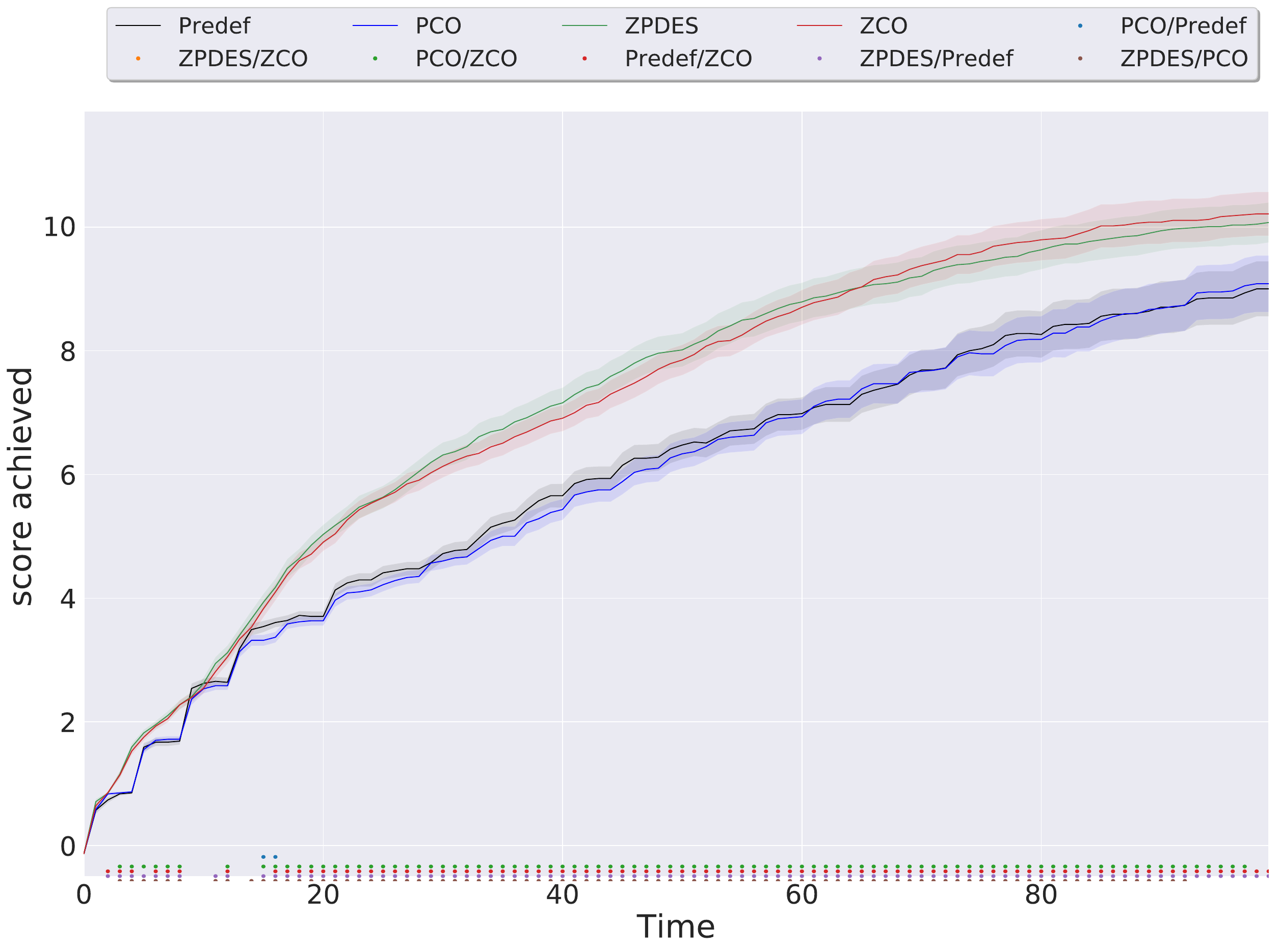}\\
   
        \end{tabular}
    \caption[Activity score over time]{After 15 steps, students working with ZCO and ZPDES are achieving and succeeding more difficult and diverse activities than student working with PCO and Predef conditions. The curves represent the average score over all students for one condition. The shaded area represent the standard error of the mean. Colored points indicate if the score differences are significant two by two for each time step through t-test procedure.} 
    \label{fig:activityScores}
\end{figure*}

It is not always trivial to grasp meaningful statistical differences between curricula over a population of student. Thereby, in order to be able to compare globally and quantitatively the curricula generated in each conditions, an ``Activity score'' has been designed (details in \ref{sec:IngameMeas}).  This score represents the level of difficulty either reached or achieved (i.e. reached \textit{and} succeeded) by the students for each type of activity. In other words, we use this score as a proxy to understand how the student activity space generally evolves across time for each conditions. 

Figure~\ref{fig:activityScores} shows the evolution of the average "Activity Score" for each condition across time. The shaded area represents the standard error of the mean and colored points indicate if  the score  differences  are  significant two by two through t-tests (as presented in the top legend of the figure). 

We can observe that before 15 steps, there is not much differences between the conditions. But, after 15 steps, the scores of the students working with ZPDES and ZCO grow faster than the scores of the students working with Predef and PCO. This means student working with ZPDES and ZCO are doing more diverse and difficult activities through time (and succeed in more diverse and difficult activities). 

\begin{figure*}[h!]
    \centering
    \hspace*{-0.8cm}
        \begin{tabular}{ccccc}
            & $t = 1$ & $t=8$ & $t=20$ &  $t = 50$  \\
            \rotatebox{90}{\hspace{0.4cm} Predef} & 
            \includegraphics[width=0.22\textwidth]{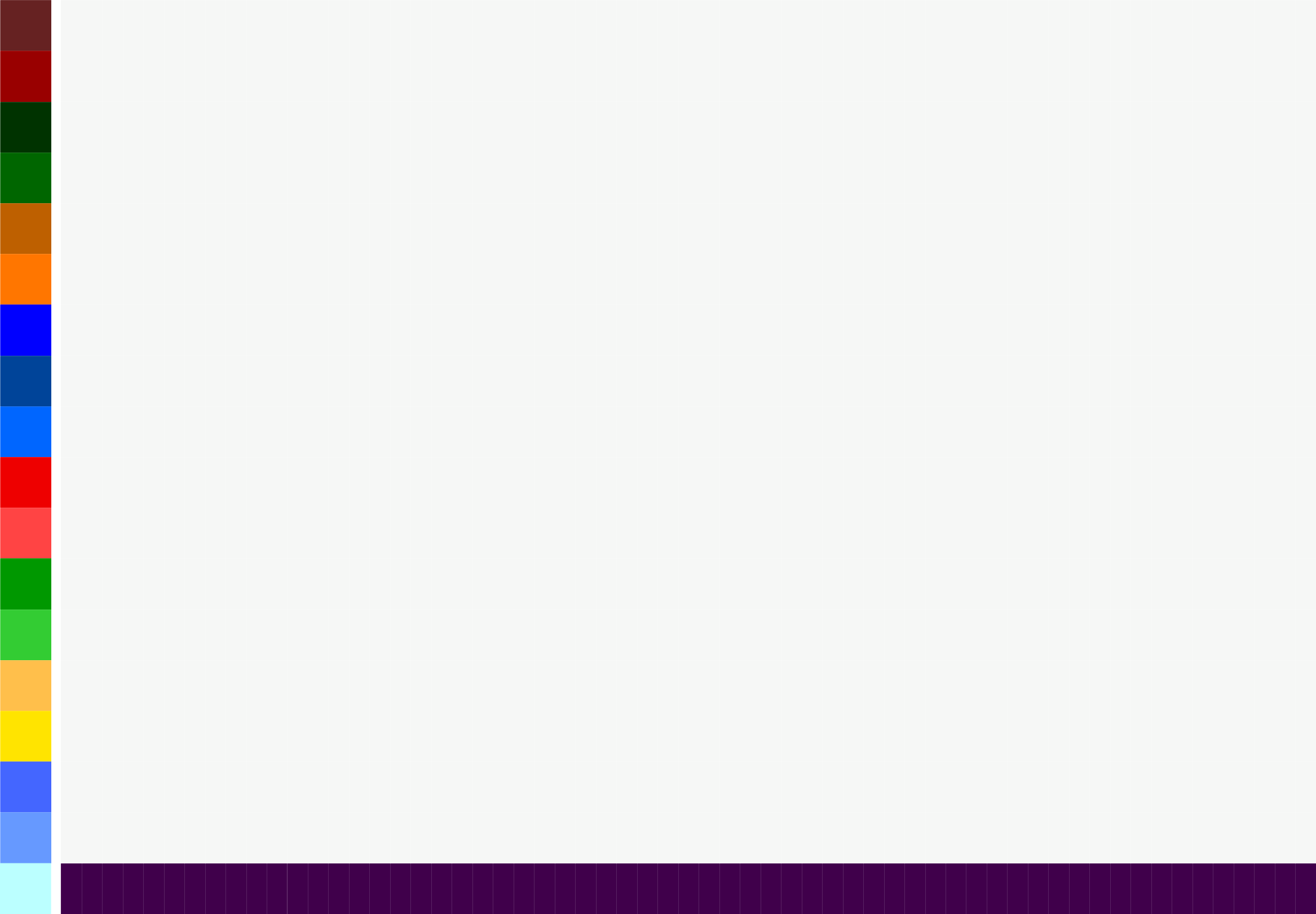} & 
            \includegraphics[width=0.22\textwidth]{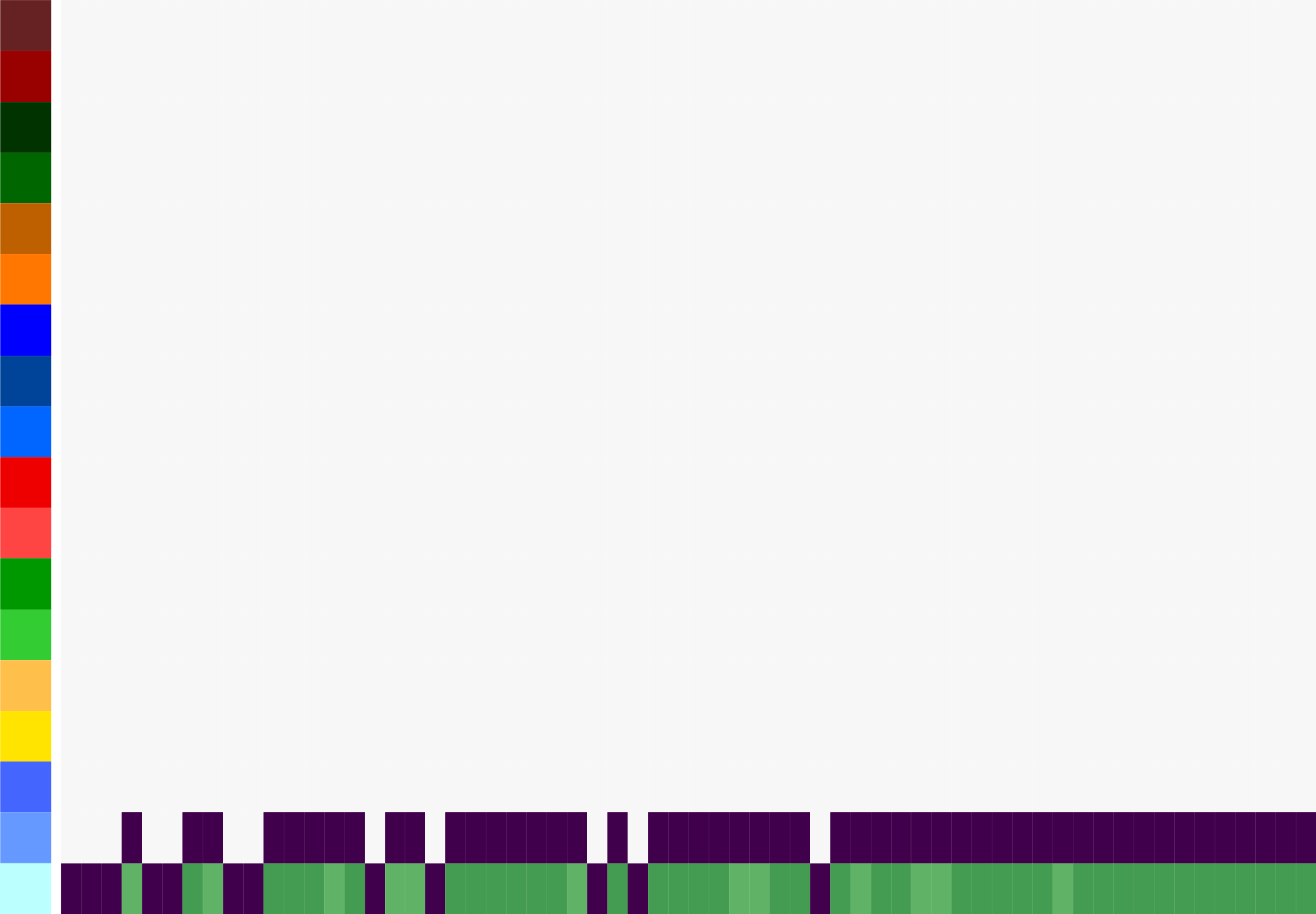} & 
            \includegraphics[width=0.22\textwidth]{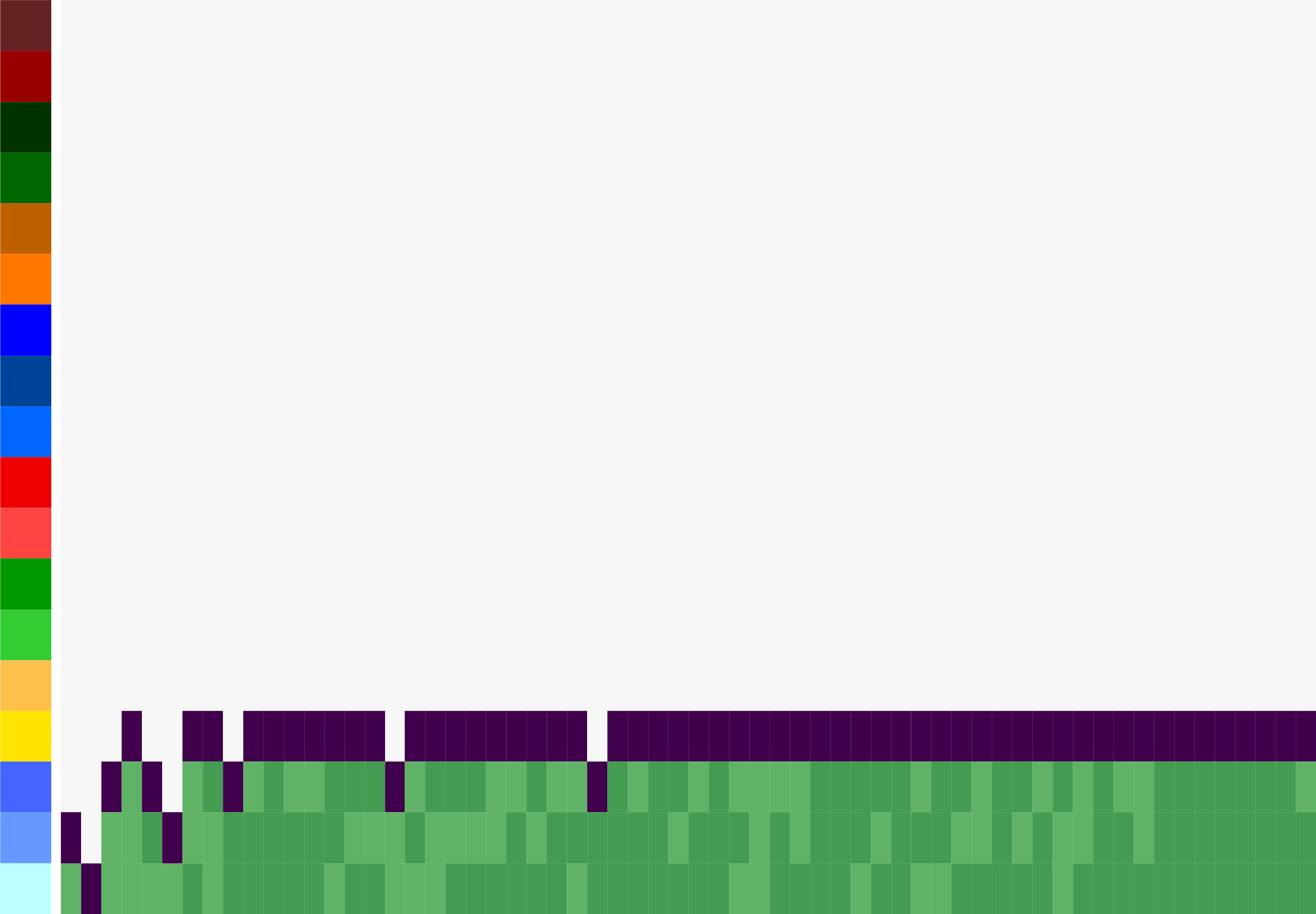} & 
            \includegraphics[width=0.22\textwidth]{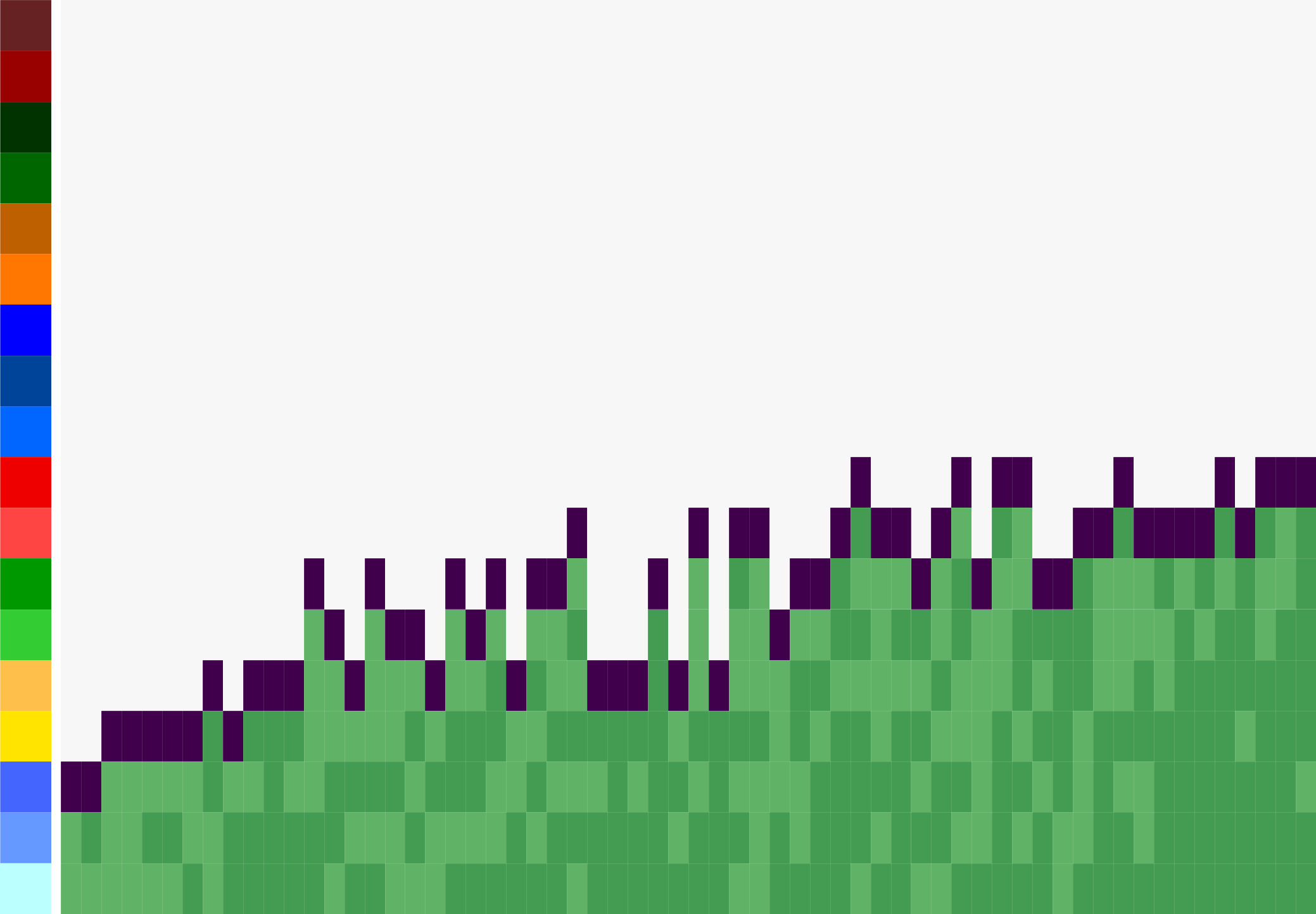} \\
            
            \rotatebox{90}{\hspace{0.4cm} ZPDES} & 
            \includegraphics[width=0.22\textwidth]{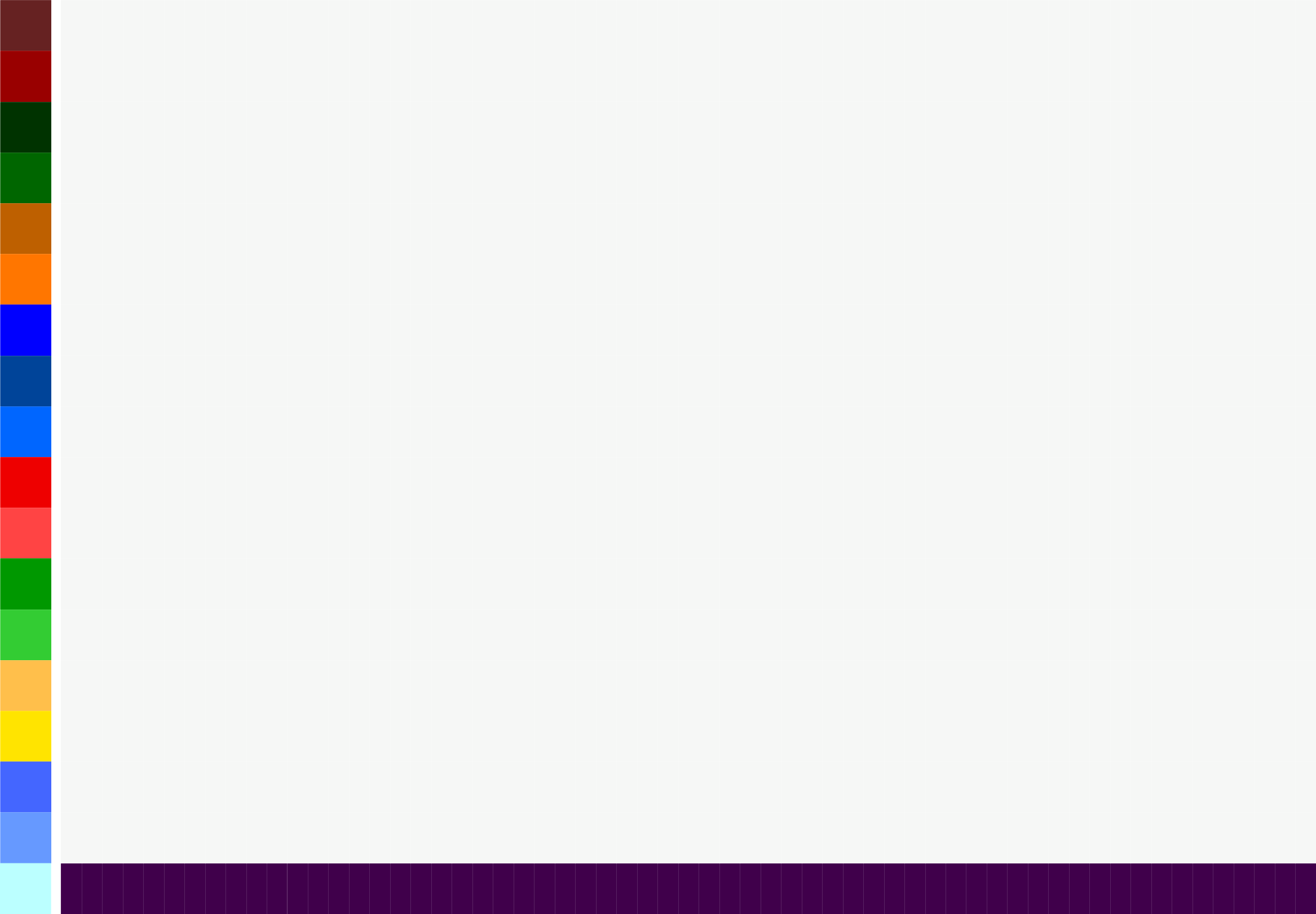} & 
            \includegraphics[width=0.22\textwidth]{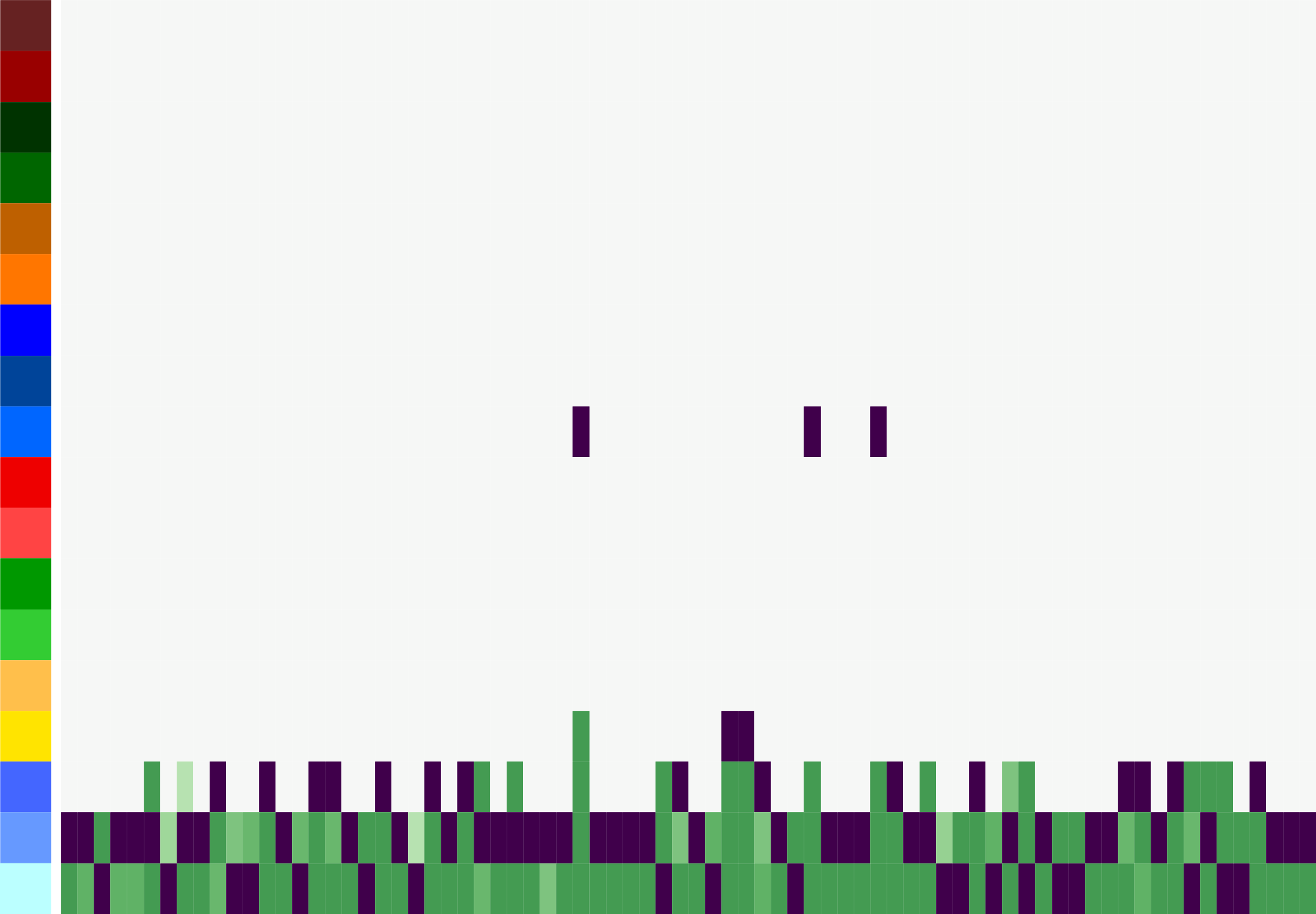} & 
            \includegraphics[width=0.22\textwidth]{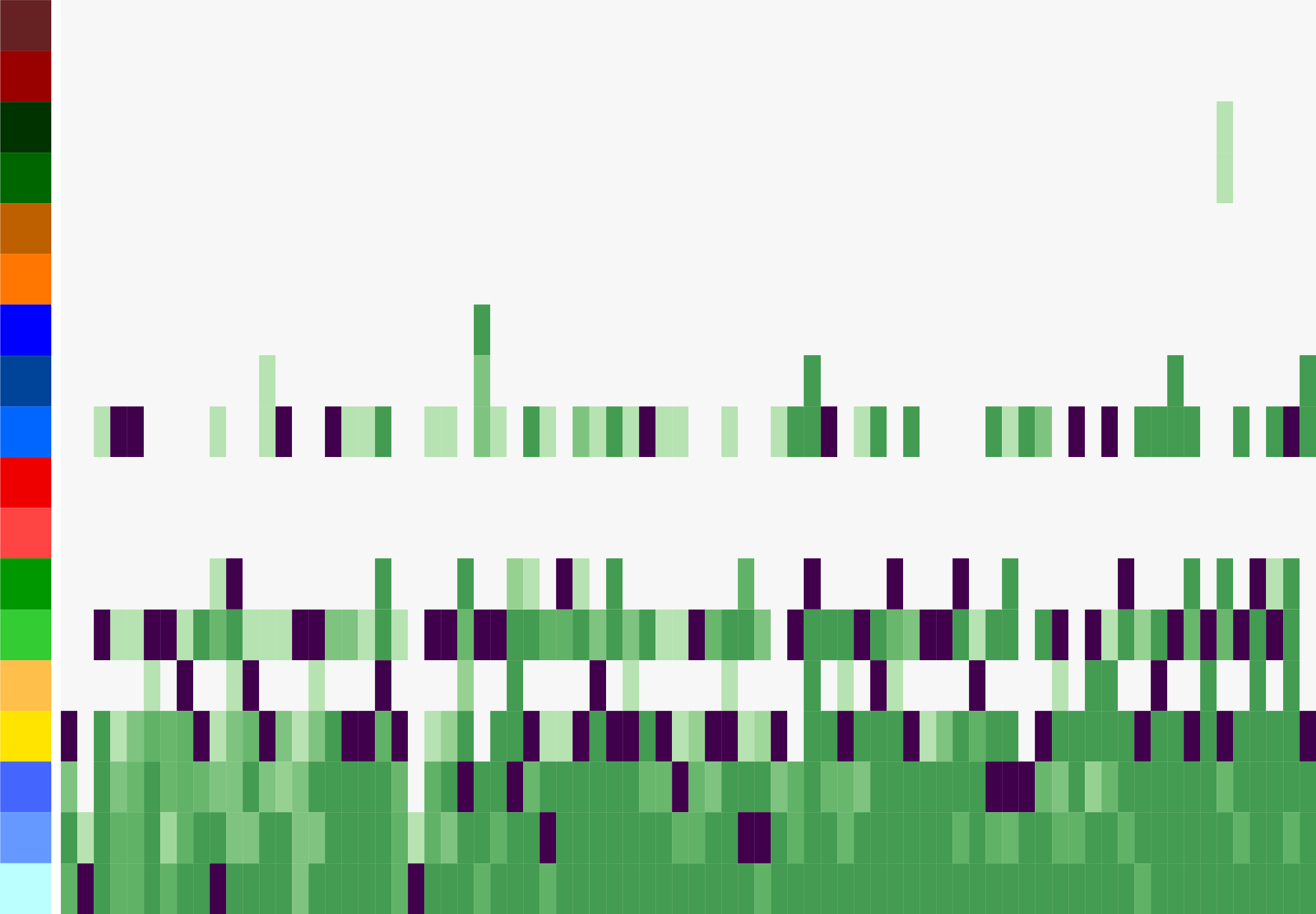} & 
            \includegraphics[width=0.22\textwidth]{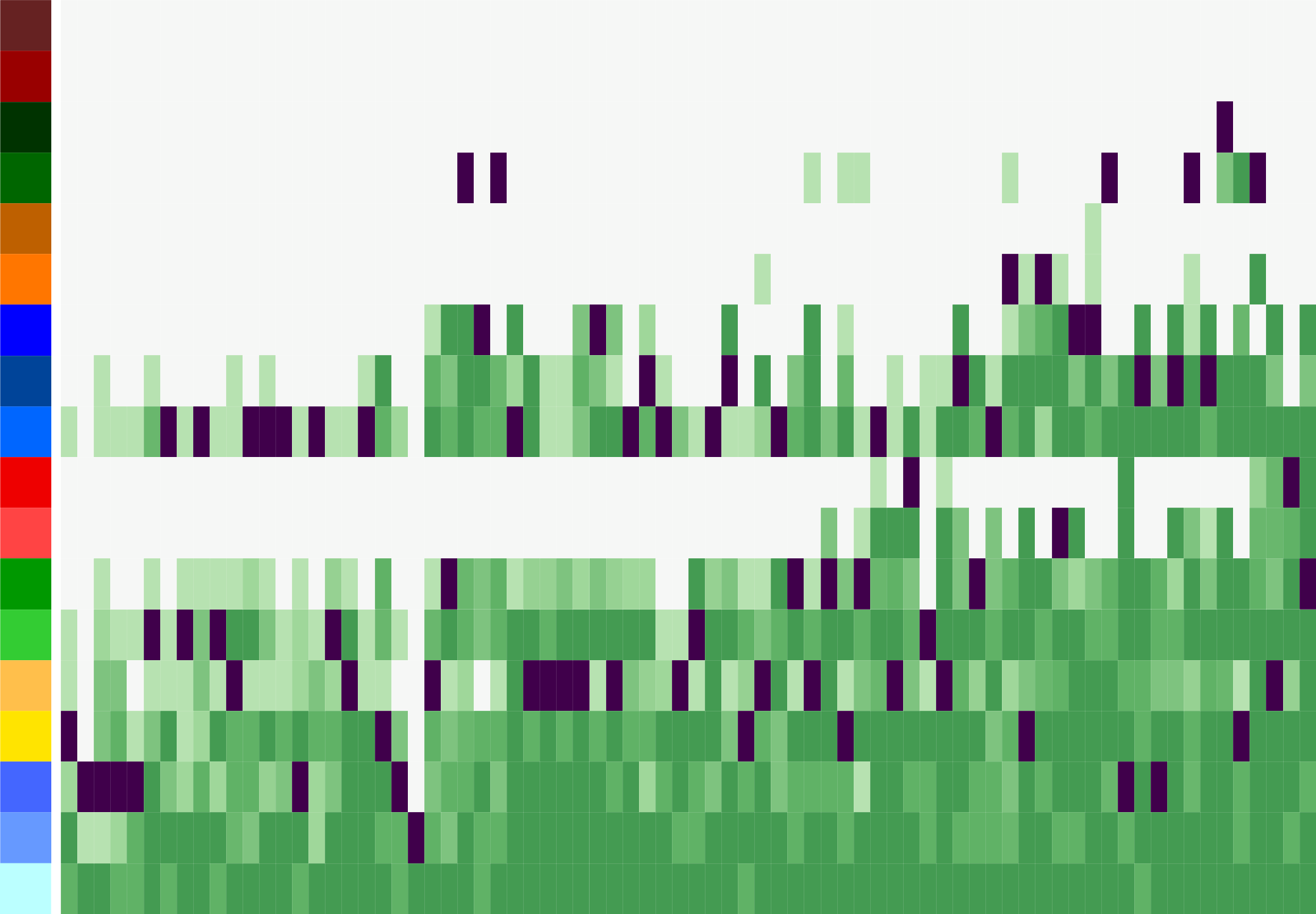} \\
           
            \rotatebox{90}{\hspace{0.4cm} PCO} & 
            \includegraphics[width=0.22\textwidth]{figures/results/chrono/file_time_all_PCO_0_rt.pdf} & 
            \includegraphics[width=0.22\textwidth]{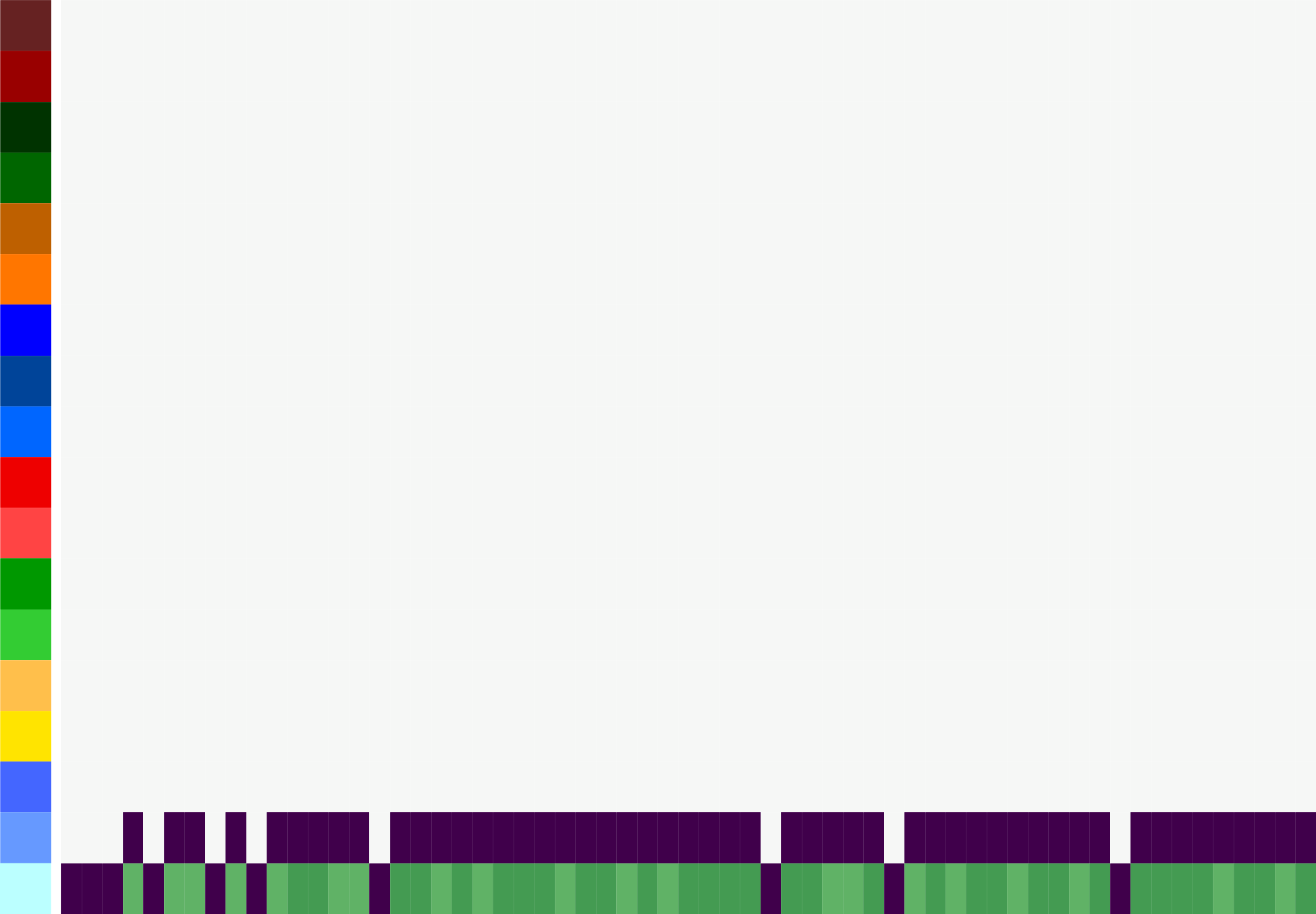} & 
            \includegraphics[width=0.22\textwidth]{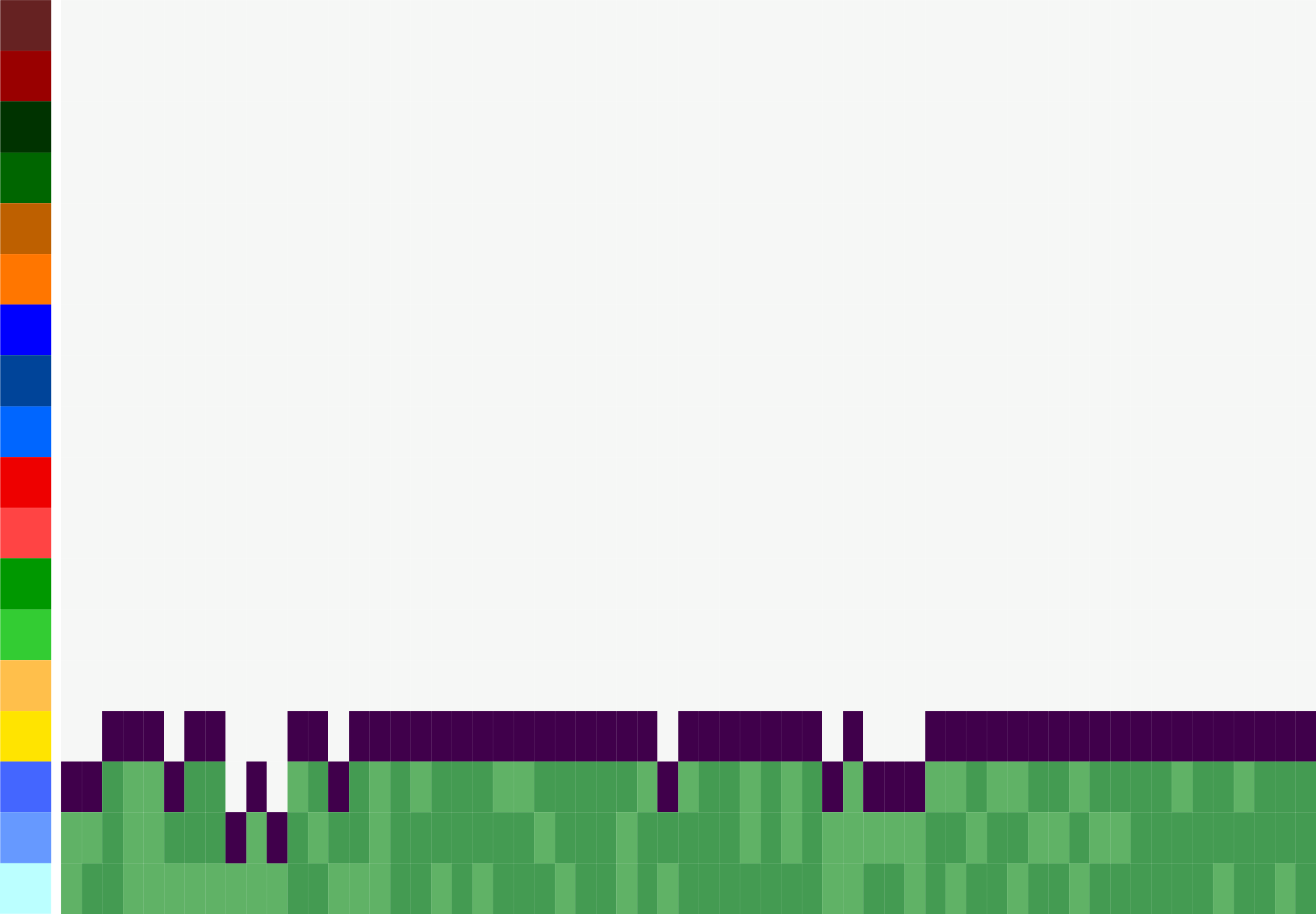} & 
            \includegraphics[width=0.22\textwidth]{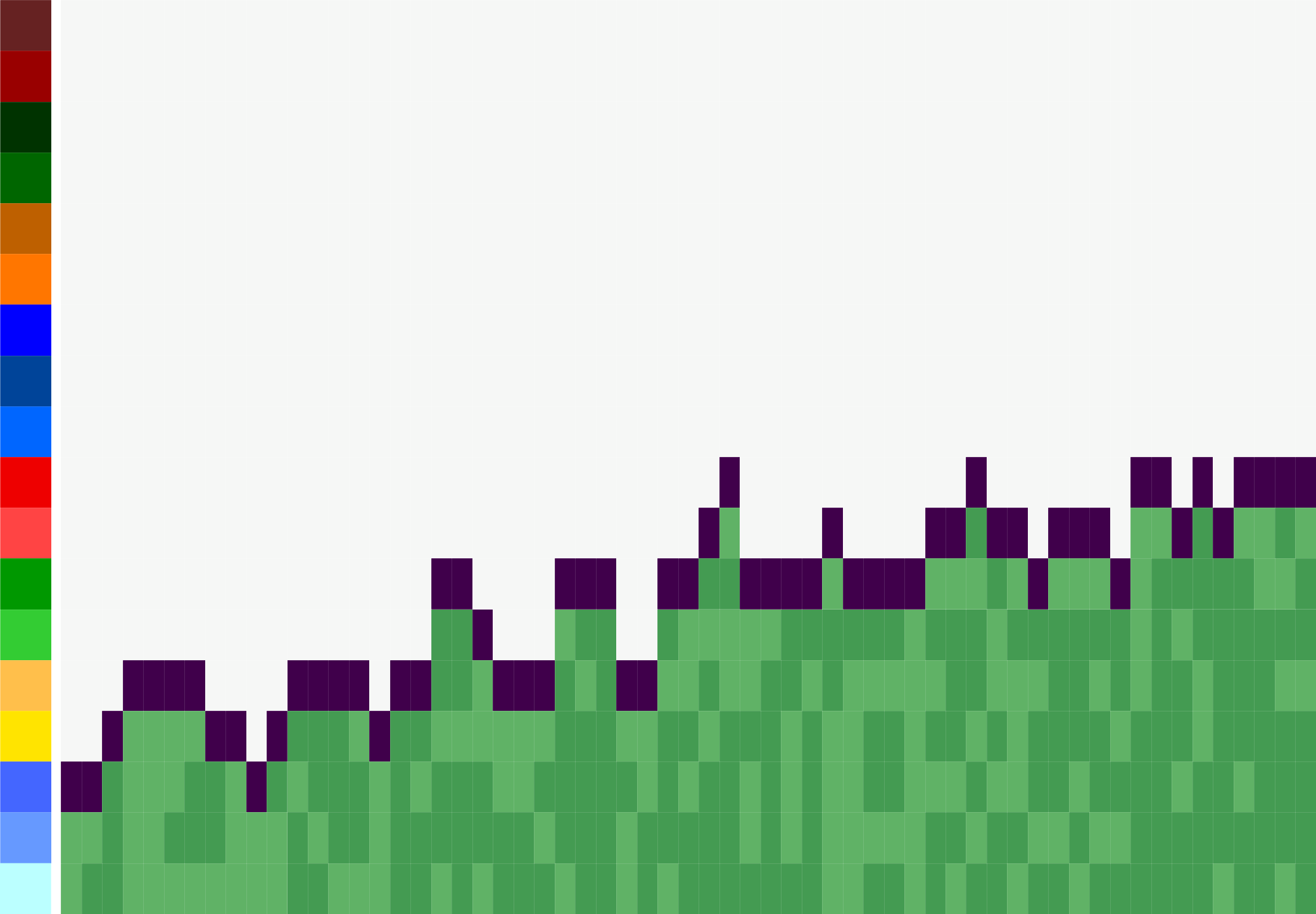} \\
            
            \rotatebox{90}{\hspace{0.4cm} ZCO} & 
            \includegraphics[width=0.22\textwidth]{figures/results/chrono/file_time_all_ZCO_0_rt.pdf} & 
            \includegraphics[width=0.22\textwidth]{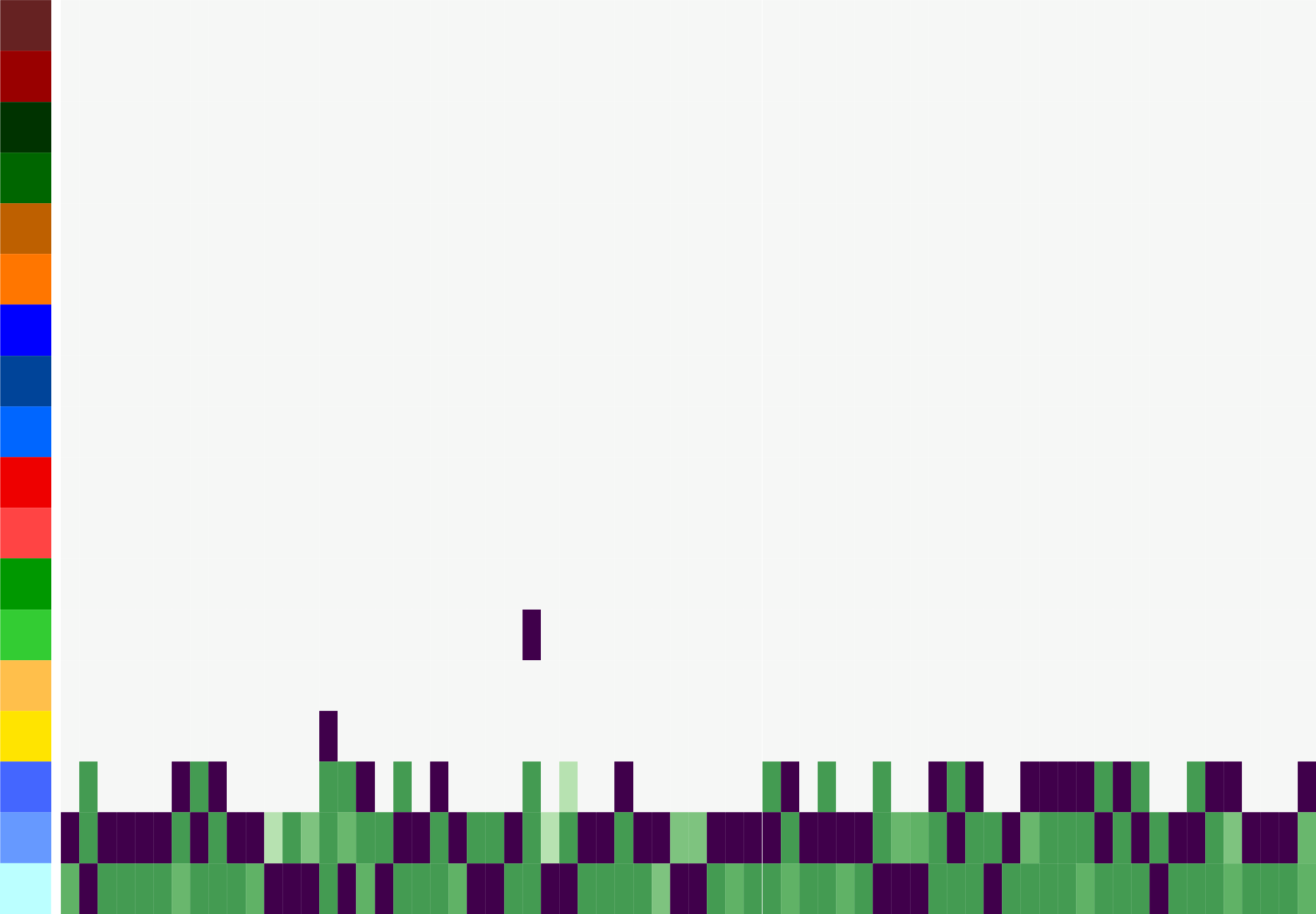} & 
            \includegraphics[width=0.22\textwidth]{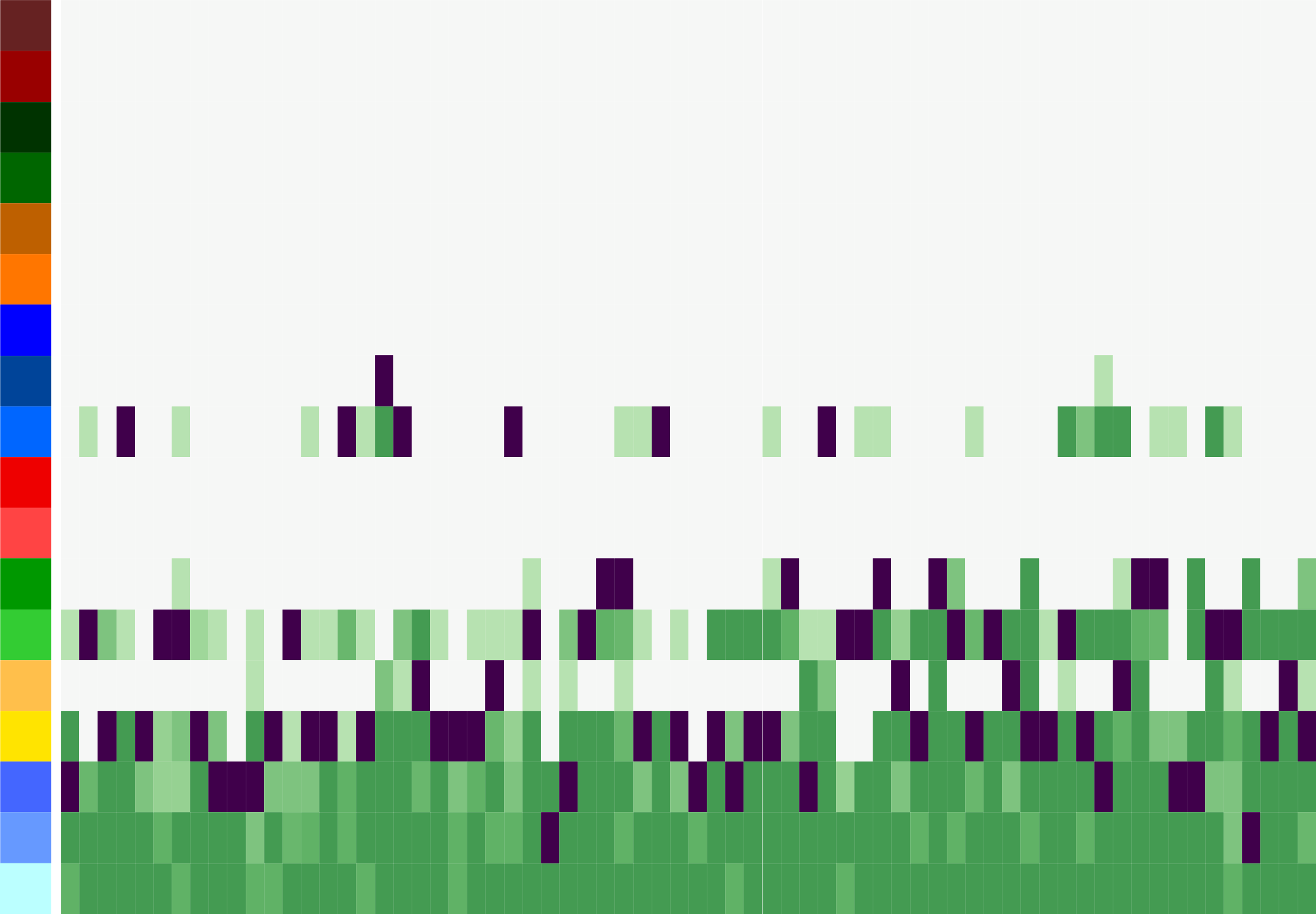} & 
            \includegraphics[width=0.22\textwidth]{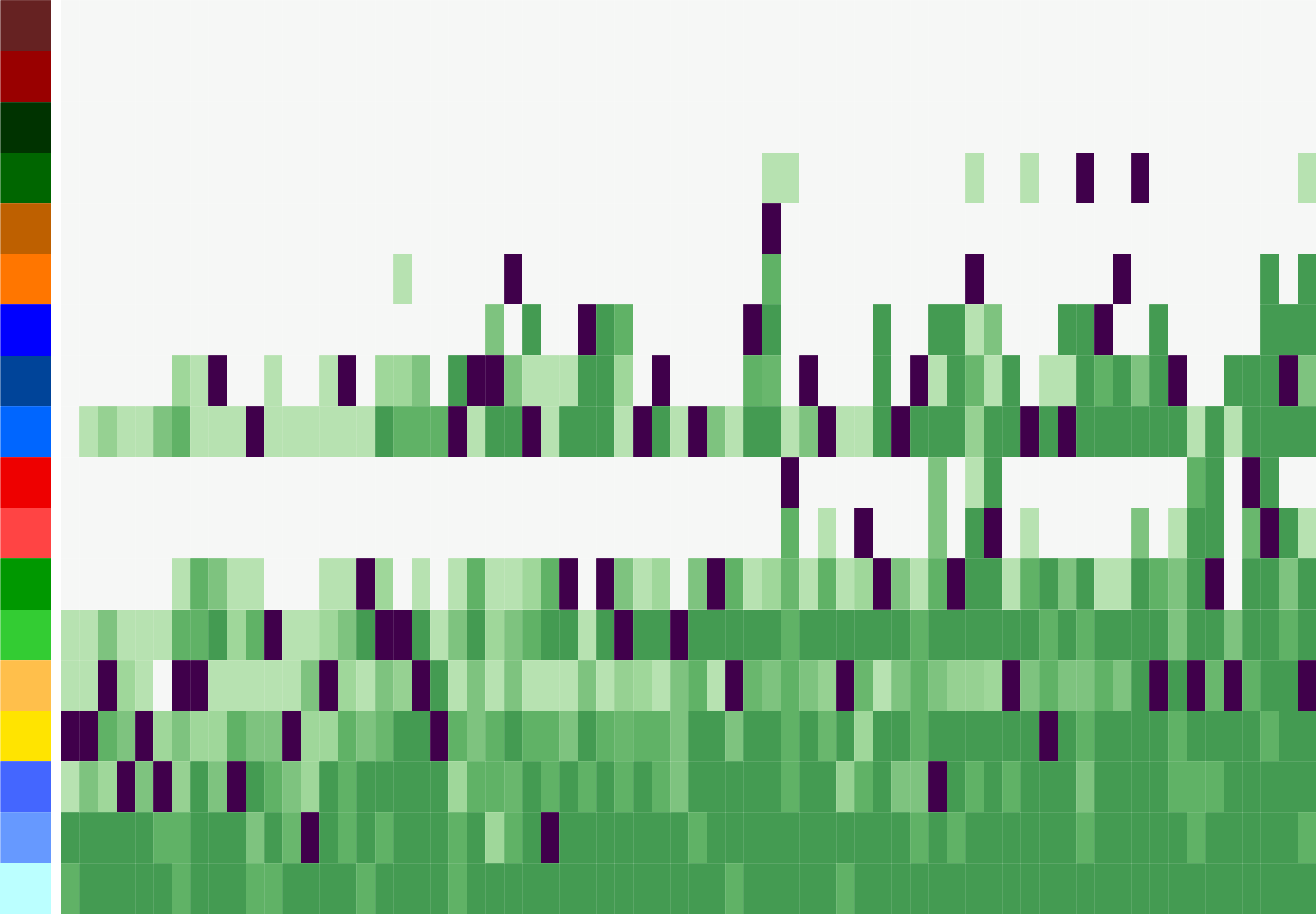} \\
            
            &{\includegraphics[width=0.22\textwidth]{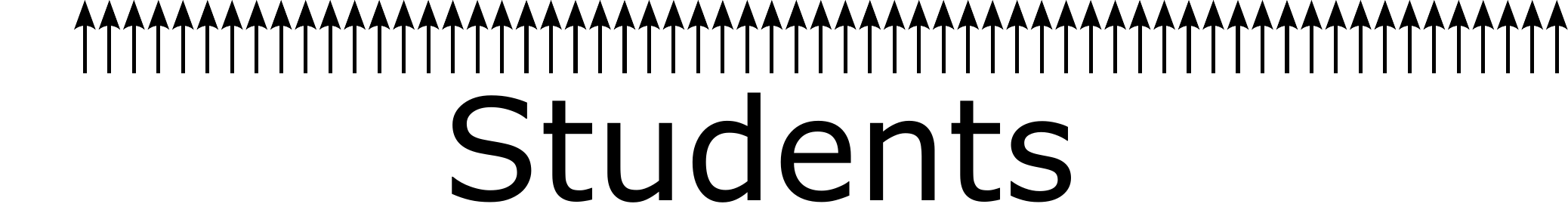}} 
            &{\includegraphics[width=0.22\textwidth]{figures/results/chrono/student_legend.pdf}} 
            &{\includegraphics[width=0.22\textwidth]{figures/results/chrono/student_legend.pdf}} 
            &{\includegraphics[width=0.22\textwidth]{figures/results/chrono/student_legend.pdf}}\\
            
            & \multicolumn{4}{c}{\includegraphics[width=0.85\textwidth]{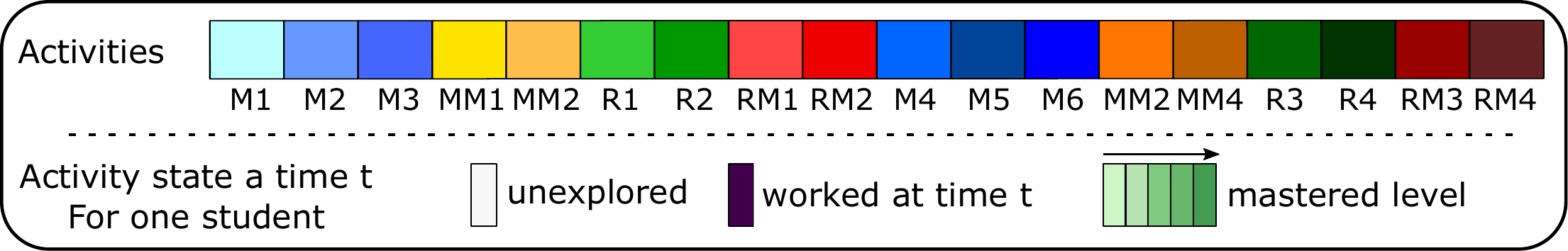}}   \\

        \end{tabular}
    \caption[Student Chronographs]{Students working with ZPDES and ZCO go through the graph of learning activities faster than students working with Predef and PCO. This way, they reach and achieve a larger set of activities. 
    There are $4$ types of activity M, MM, R and RM with their related levels (M: $6$, MM: $4$, R: $4$, RM: $4$). They are ordered here in a colored band displaying a relative difficulty hierarchy to be able to facilitate the visualisation of the students' evolution across activities. Each cells represent the state of an activity for a student at time "t"; white sells for not explored, purple for activity done at time "t"; green for explored activity  }
    \label{fig:magHistoExp3TimeAct}
\end{figure*}

Also, giving a contextual choice does not seem to affect the activities done by the students due to the fact that ZPDES and ZCO scores are really similar, as it is for Predef and PCO. 
To have a more qualitative overview of the student activities, figure~\ref{fig:magHistoExp3TimeAct} shows the curriculum of each student through the activities made at $4$ different times steps. For a given time step $t$ and condition, a matrix slot represents the state of an activity (ordinate) for a particular student (abscissa). A slot is grey if a student has never explored the corresponding activity and it is purple if the student is doing this activity at time $t$. When a student has explored an activity, the slot is tinted green depending on the student's success rate (light green: low, dark green: high). 

Across time, this figure allows to confirm the observation made previously which is that student working with ZCO and ZPDES are able to explore a larger set of activities than the ones working with PCO and Predef. We can also confirm that contextual choice does not affect the overall profile of activities proposed by ZPDES and Predef.   

\subsection{How does LP-based personalization impacts learning effectiness as compared to hand-designed curricula?}
\label{sec:resLearning}

The learning effectiveness of each condition is evaluated through comparison between pre- and post-test results. The pre- and post-test 
are composed of 20 items scoring from 0 to 1 (max score is 20). The pre-test happens at the beginning of the first session, while the post-test happens at the end of the last session. The pre- and post-test are presented on the tablet on a dedicated interface (different from the ITS one). Each item of the test evaluates the student over knowledge and skills related to money manipulation, number composition, addition or subtraction (similar to the skills and knowledge trained in the ITS). Both tests include the same items organised in the same order but the items' wording have randomly selected values for each item and each student (with verification that no items in the post-test have the same values in their wording as the ones in the pre-test for one student).  

The statistical procedure used in this section consists of three-way mixed ANOVA (algo x choice x  pre/post) on the Math-tests score (pre- and post-measurement of student performance), with the pre/post factor as within-subject factor.
The algorithm factor includes the two conditions (ZPDES or Predef). And the choice factor include also two conditions (with or without choice). The p-value threshold is $\alpha=0.05$. Pairwise comparisons are carried out with the Least Significant Difference (LSD) and Bonferroni procedure for corrected comparisons.

\paragraph{What is the impact of LP-based personalization without choice?}

\begin{figure*}
 \centering
     \begin{tabular}{cc}
        Pre-Test & Learning Score \\
        \includegraphics[width=0.49\linewidth]{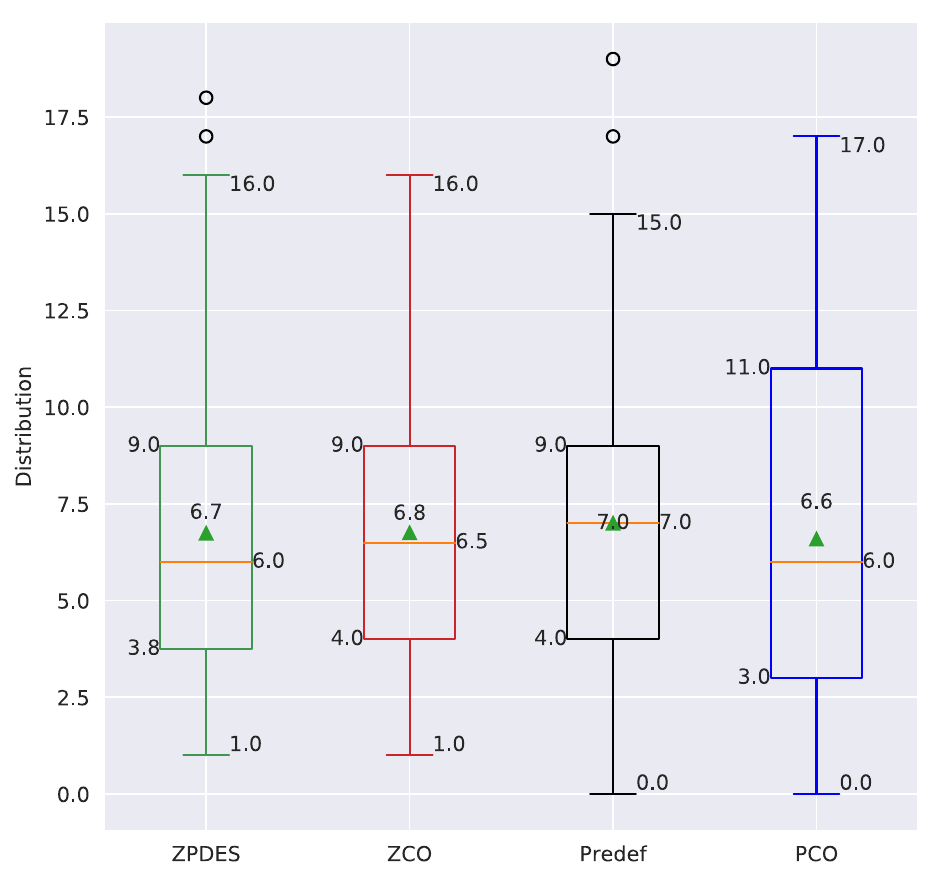}&
        \includegraphics[width=0.49\linewidth]{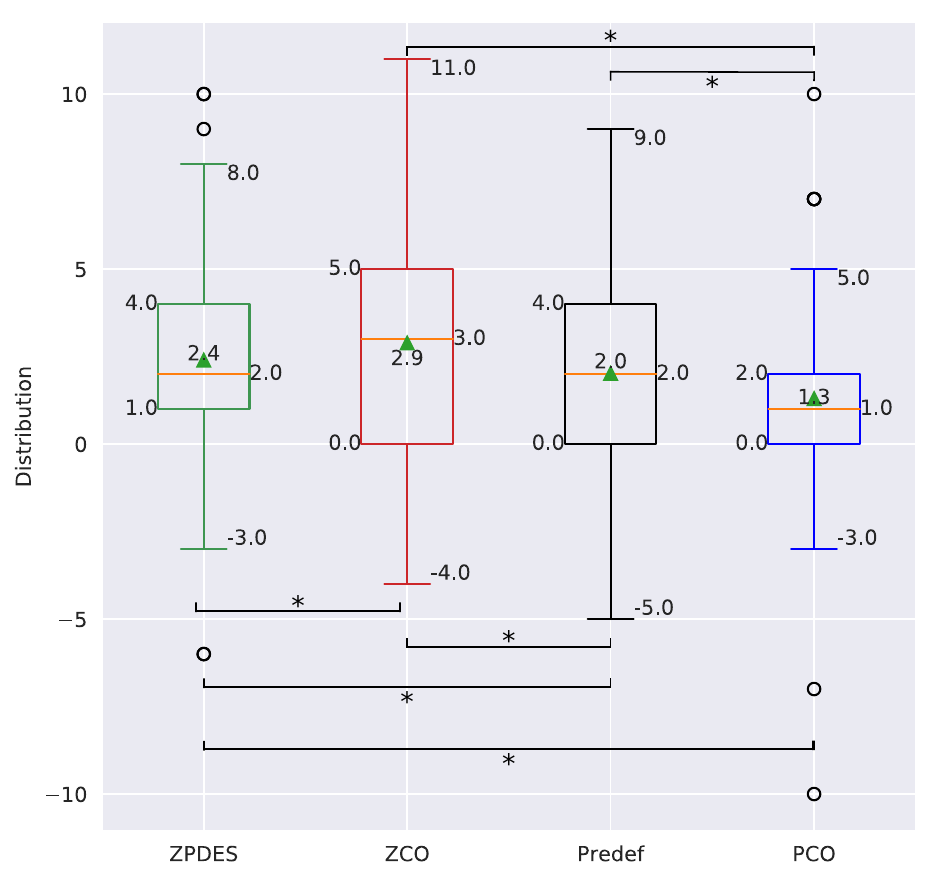}
     \end{tabular}
     
     \caption{Boxplots presenting the Pre-test scores and the Learning score, i.e the difference between Post-test score and Pre-test score for the four conditions. The Pre-test scores are homogeneous among the populations, assuring a fair comparison. The Learning scores are ordered as follow ZCO $>$ ZPDES $>$ Predef $>$ PCO, giving the order in term of learning efficiency between each condition.}
    \label{fig:diff}
 
 \end{figure*}

The main significant effect revealed an increase of the test score across time (pre/post factor effect, $[F(1,261)=129.25$, $p-value=0.000$, $\eta^2 = 0.331]$) which is boosted under the ZPDES condition compared to Predefined condition (algo x pre/post  effect, $[F(1,261)=40.076$, $p-value=0.003$, $\eta^2=0.034]$). This effect combined with the examination of the marginal means (Predef: pre/post $mean=6.83 (sd: 0.353)$ / $8.36(sd: 0.396)$, ZPDES: pre/post $mean=6.74 (sd: 0.344)$ / $9.38 (sd: 0.363)$) shows that children working with ZPDES algorithm learned more than the ones working with the Predefined sequence algorithm (visual support on Fig.~\ref{fig:diff}).

\paragraph{Does the possibility to express choice boost learning?}

Even more interestingly, the three-way interaction is significant (algo x choice x pre/post effect $[F(1,261)=17.319$, $p-value=0.049$, $\eta^2=0.015]$), the learning benefit from ZPDES condition is increased by the choice condition, whereas we observe the opposite for Predefined condition (Detrimental effect of choice). Pairwise comparisons indicate  significant differences between PCO and ZCO for the post test score according to LSD procedure ($p-value = 0.014$), and only marginal differences according to Bonferroni procedure ($p-value = 0.08$).

\subsection{How does LP-based personalization impact the emotional valence of learning experience and learner's motivation as compared to hand-designed curricula?}
\label{sec:res_emoValence}

The emotional valence of the learning experience is assessed by an emotional scale. Thought this scale, the student can express how (s)he feels by moving a cursor from the best moment of his life to the worst one. The students answer this scale 3 times during each session (start, middle and end). 

The statistical procedure used here consists of a two-way ANOVA (algo x choice) on the Emotional Scale Score (summative score of emotional valence of learning experience collected during each Kidlearn session, see Kidlearn-related learning experience section).

The algorithm factor includes the two conditions (ZPDES or Predef). And the choice factor include also two conditions (with or without choice). The p-value threshold is $\alpha=0.05$. Pairwise comparisons are carried out with the Least Significant Difference (LSD) and Bonferroni procedure for corrected comparisons.

\begin{figure*}
 \centering
 \begin{tabular}{cc}
    EmoScale & Motivation Score  \\
    \includegraphics[width=0.49\linewidth]{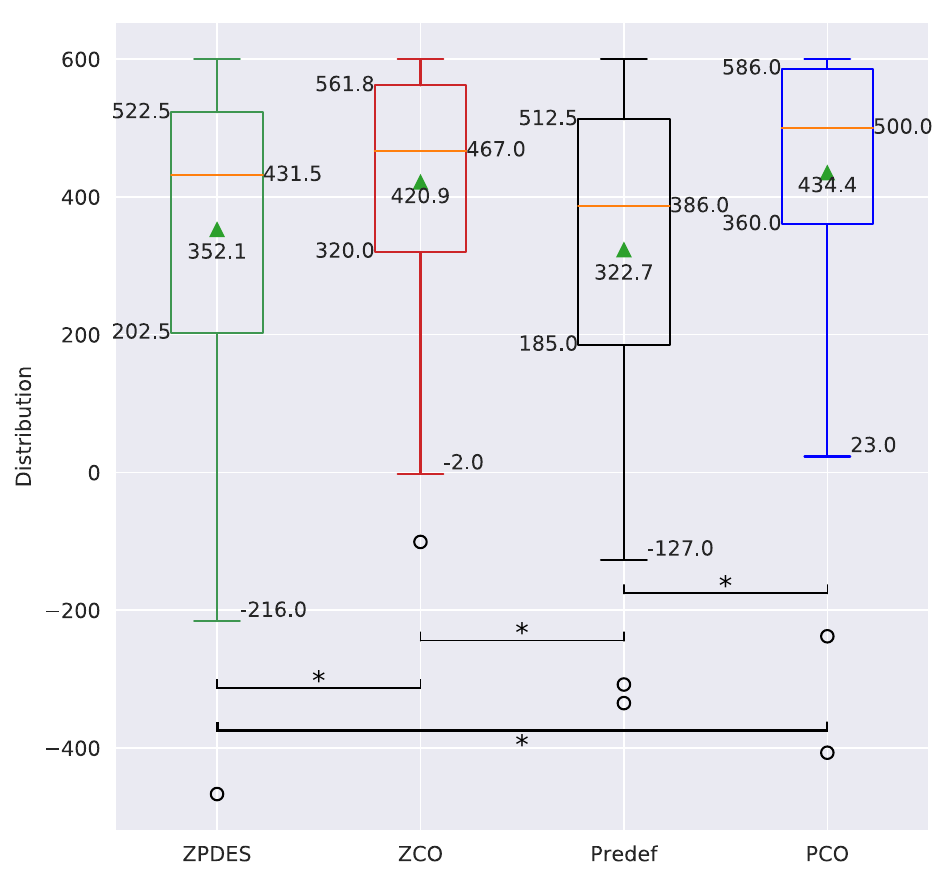} &
    \includegraphics[width=0.49\linewidth]{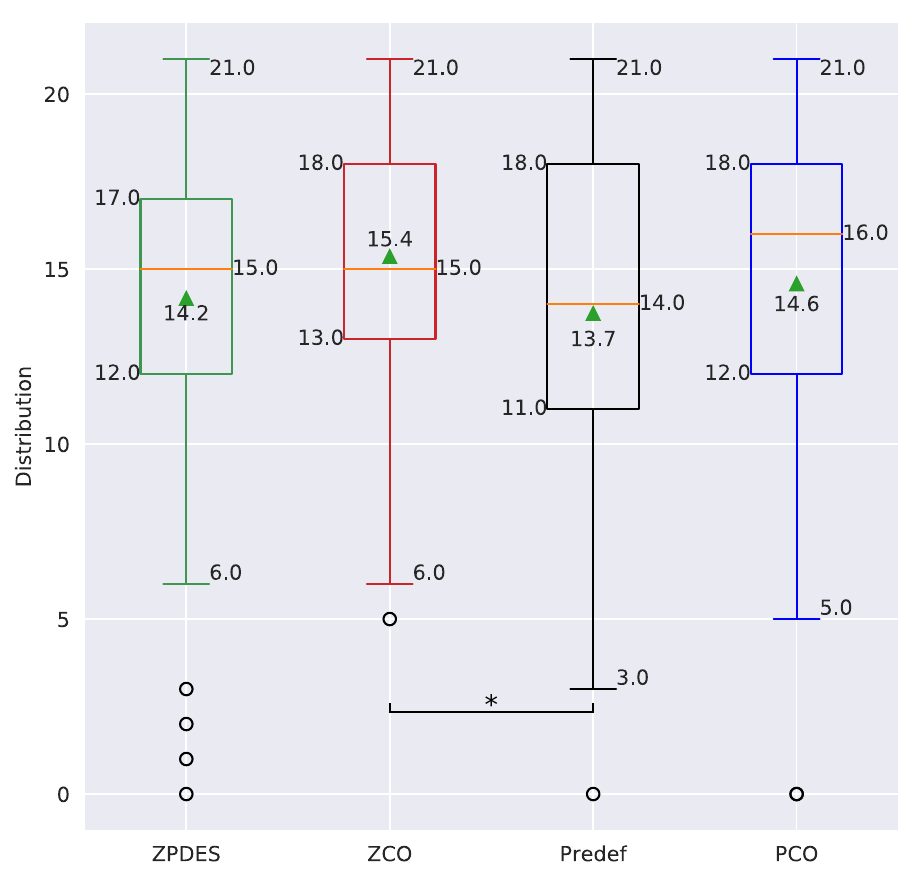}
 \end{tabular}
    
     \caption[Motivation Boxplot]{Boxplots presenting the Emotional Scale score on the left and the Motivation score on the right. Students working with ZCO and PCO show the highest EmoScale scores while students working with ZCO show the highest Motivation score, followed by PCO ad ZPDES and Predef present the lowest score. }
    \label{fig:res_mot}
 
 \end{figure*}

The main significant effect revealed a difference for the Emotional Scale score between students who have choices and student without choice (Choice, $[F(1,261)=12.060$, $p-value=0.001$, $\eta^2=0.044]$). This effect combined with the examination of marginal means, (Choice: EmoScale $mean=426.05 (sd: 183.58)$, No Choice: EmoScale $mean=338.86 (sd: 226.96)$) shows that children working with the possibility to choose the object of the exercise feel better than the ones who does not have the possibility to choose, which suggests they are more satisfied of their leaning experience (visual support on Fig.~\ref{fig:res_mot}). 

\paragraph{Does the possibility to express choice boost motivation?}

The motivation is evaluated through Vallerand’s questionnaire\cite{vallerand1992academic, vallerand1989construction}. It is based on Self-Determination Theory\cite{deci1985general} and is commonly used to assess the elicitation of intrinsic and extrinsic motivation (e.g. \cite{desrochers2006elaboration}). It is composed of 21 items about the student's experience during the experiment sessions.

We also conducted a two-way  ANOVA (algo x choice) on the Motivation score. The algorithm factor includes the two conditions (ZPDES or Predef). And the choice factor includes also two conditions (with or without choice). The p-value threshold is $\alpha=0.05$. Pairwise comparisons are carried out with the Least Significant Difference (LSD) and Bonferroni procedure for corrected comparisons.

There is no significant effect but we can observe a tendency showing a difference between students who have choices and student without choice  (Choice, $[F(1,261)=3.449$, $p-value=0.064$, $\eta^2=0.013]$). From this tendency, the examination of pairwise comparisons reveals a significant difference between ZCO and Predef according to LSD procedure ($p-value=0.034$), but not from Bonferroni procedure ($p-value=0.205$). This tendency combined to the examination of margin means (ZCO: MS $mean=15.353 (sd:0.528)$, Predef: MS $mean=13.726 (sd:0.553)$)  seems to support that giving choice allows students to have a more motivating experience (visual support on Fig.~\ref{fig:res_mot}). 

This fits with the greater positive emotional experience elicited by the choice condition, and more particularly for ZCO condition.

\subsection{Does a positive relation exist between LP-based personalization and subsequent learning performance and motivation~?}
\label{sec:perfMotRelation}

In order to establish a relationship between the learning effectiveness and the learning experience according to each experimental condition, correlations for each experimental condition (Predefined, PCO, ZPDES, and ZCO) were made between the following 3 measures (see Tab.~\ref{tab:inter-correlation})  : 1) Learning score (difference between pre-and post-test) ; 2) Kidlearn progression (final activity score defined in \ref{sec:IngameMeas}) ; 3) Motivation score (defined in \ref{sec:motivMeasure}).

Importantly, the ZCO condition is the only condition where it is possible to observe positive relations between the learning score and the learning experience with the ITS in terms of both kidlearn progression and the motivation state (respectively $r=.21$ and $r=.27$). This observation correlates with the LP hypothesis that effective learning and intrinsic motivation are linked to activities in which learners progress and can exercise self-determination. Similarly, ZPDES condition induces a positive relationship between learning score and Kidlearn progression ($r=.32$). Taken together, these observed correlations strongly support the link between the progression in the Kidlearn app, enabled by this personalizing algorithm, and the actual learning progress. 

In contrast, for the PCO condition, no correlation is significant (see table~\ref{tab:inter-correlation}). This suggests that there is no link between the Kidlearn progression and the level of motivation elicited by the choice or the actual learning progress. As already mentioned, the children tended to be motivated by the choice opportunity but this motivation was not sufficient to actually progress while working with the predefined sequence. 
\begin{table}[ht!]
    \centering
   
    \begin{tabular}{ccc|c}
        & & \multicolumn{2}{c}{Kidlearn-related learning experience} \\\hline
        \multicolumn{2}{l}{Learning score} & Kidlearn progression &  Motivation score \\
        \multicolumn{2}{l}{(Pre/post difference)} & (Final activity score) & \\\hline
        
        Predef  & R value & $.001$ & $ \bf .29^+$  \\
                & P value & $ ns.$ & $ .02$  \\\hline
        PCO     & R value & -$.006 $ & $ .11 $ \\
                & P value & $ ns. $ & $ ns. $  \\\hline
        ZPDES   & R value & $ \bf .32^+ $ & $ .07 $ \\
                & P value & $ .005 $ & $ ns. $  \\\hline
        ZCO     & R value & $ \bf .21^+ $ & $ \bf .27^+ $   \\
                & P value & $ .07 $ & $ .02 $  \\\hline
                
    \end{tabular}
    \caption{Bravais-Pearson inter-correlation between Learning score (pre/post difference) and the Kidlearn experience scores with the ITS application (Kidlearn progression and the motivation score).  
Notes. ns. = non significant. According to Fisher’s transformation procedure (with the limit values for Z at 1.96), r  values comparisons revealed no significant difference across conditions }
    \label{tab:inter-correlation}
\end{table}

Finally, for the Predef condition, no relation is observed between the learning score and the Kidlearn progression while the learning score under this condition is positively related to the student motivation ($r=.29$). As this condition induces the lowest learning score and the lowest motivation scores in children, this last correlation hints that learning outcome from Predef condition may be mainly related to the student's prior motivation, where the most motivated children do best, and the least motivated do worst, thus widening the differences in learning in this condition between the most motivated and the least motivated. This interpretation is corroborated by the much larger range of performances for these two variables in the predefined condition compared to the other three conditions. 

Overall, this means that the ZCO is the best learning condition yielding actual learning progress associated to the learner's motivation. Additionally, only the LP-based conditions (ZPDES and ZCO) yield a reliable relationship between the  progression across ITS based intervention and the real outcome in terms of learning benefit.
In other words, the positive relation between LP and motivation is boosted when students can exercise their self-determination through choice in the ZCO condition. This solidifies the hypothesis that LP is correlated with/generates intrinsic motivation only when learner has the ability to choose, i.e. feels autonomous.

\subsection{Is the impact of LP-based personalization modulated by individual characteristics of learners~?}
\label{sec:modulateInd}
All the previous analyses have in addition been conducted with ANCOVA analyses where the covariable was related to individual characteristics. Particularly we investigated the mediating effect of school satisfaction, digital technology experience, gender and age (see Profil metrics in \ref{sec:profileMeasure}).

No significant results have been observed revealing that the present results are robust to mediating effects related to the studied individual factors (see Tab.~\ref{tab:fraq_stats} in \ref{sec:appendIndCharac}).

\subsection{Synthesis}

From the overall data, we can infer there is a double beneficial effect to the combination of ZPDES and choice on the learning and the motivational levels. This corroborates the results describing choice effect as a positive lever on motivation and performance \cite{leotti2011inherent, murayama2013self, cordova1996intrinsic}. 
However, our results show that the positive effect of choice, in terms of learning effectiveness, is algorithm-dependent. The choice  is beneficial for ZPDES algorithm, whereas it is detrimental for predefined algorithm. This can be interpreted as, without a relevant teaching strategy, the choice will act as a distractor and students will focus more on the choice and less on the activity. In other words, allowing choice in inappropriate teaching strategies is deleterious for the students' learning, although they enjoy to make choices.

%% file: discuss.tex

\section{Discussion}

On a large sample of students, our results clearly indicate that personalization of the learning path via an algorithm that estimates the proximal learning zone by maximizing LPs is more effective in terms of learning outcomes than "linear design" strategies that only adapt the pace and number of exercises across the linear curriculum.  This result is consistent with a systematic review on ITS (\cite{ma2014intelligent}), indicating that personalization is more effective than one-size-fits-all instructional design and that ITS are more effective than traditional whole-class instructional methods. 

Specifically, for the first time, we report an extensive study showing that personalizing the pathway according to the student's LP improves learning performance while producing a positive and motivating learning experience, regardless of several learner characteristics such as gender, experiences with technology, past experiences with the activity being trained, or the student's experiences and perceptions of school. Taken together, this empirically supports the robustness of LP-based personalization to diverse student characteristics (known to be critical to learning). Also, in line with our hypotheses, we show for the first time also the added value for learning outcomes of associating the LP-based individualization of the learning path with a playful feature allowing self-determined decisions, yielding a synergy of intrinsic motivations elicited by both the LP (as assumed in the LP hypothesis, \cite{oudeyer2007intrinsic2, lopes2012exploration, colas2019curious,portelas2020automatic}) and by “gamification” strategies (e.g.,\cite{proulx2017learning,tyack2020self}). 

It is noteworthy that this positive synergy on the learning outcome is associated with a positive and motivated learning experience. As a result, in the ZCO condition, instructional effectiveness (pre-/post difference) is also positively correlated with motivation scores. 

Conversely, we show a deleterious effect of the association of a playful feature with a linear learning pathway in terms of real learning outcome contrasting with a positive and motivated learning experience for the students. As a result, the correlation between learning score (pedagogical effectiveness) and intrinsic motivation was not significant. This result is particularly insightful because it highlights that a positive and motivated learning experience via a "gamification" strategy that elicits intrinsic motivations, is not sufficient to improve learning outcome. In other words, in that case the attention-grabbing power of games can lead to a distraction from the pedagogical objectives of the activity. 

This detrimental effect of "gamification" on learning performance under the PCO condition mirrors findings in children about the motivational conflict between immediate and delayed rewards (also called  want-should conflicts (\cite{bitterly2014dueling, grund2015torn, bernecker2021no}). In our case of PCO condition,  the choice of object for each exercise can be seen as an immediate reward (without learning gain expected) while the learning progression into the kidlearn are delayed rewards not very attractive due to their small magnitude related to the "one-size-fits-all" design of this condition. Overall, this reversal effect of "gamification" for linear learning path invites to be cautious when using “gamification” strategies for teaching purposes. Today, one of the great challenges of modern education is that of capturing the attention of students and creating engagement for learning tasks. In light of our results, using “gamification” strategies to enhance motivation and learning is effective only if ITS features actually fosters learners’ learning progress as provided by our LP-personalization. Such a result is consistent with the self-determination theory applied to education stressing intrinsically motivated learning for really meeting learners’ autonomy and competence needs (\cite{ryan2020intrinsic, alamri2020using}). 

Finally, a very salient result to highlight are the correlations observed between the learning score (pre-/post difference, i.e. progress) and the learning progression within the Kidlearn ITS.  Positive relationships are observed for the two conditions with LP-based personalization, but not for the two conditions with linear pathways. Hence, the learning progress observed post-intervention is really linked to the learning progress obtained through LP-based personalization. In contrast, such an assertion is not possible for the two conditions with linear pathways since the correlations are not significant. Indeed, the linear pathway with gamification seems to have made children less focused on the learning task as explained above and the linear path alone seems to be the less motivating, thus, the learning progression in this conditions seems to  reflect a combination of prior level with very little effective learning from activities and a little bit of learning from test-retest learning effect (\cite{roediger2006power}). 
Consequently, it can be argued that tools for visualizing learning paths with ZPDES or ZCO, as well as giving feedback to the student, or an instructional monitoring interface for teachers, has the potential to give reliable hints on the student's learning progress . 


\section{Conclusion}

The present field study assessing our ITS approach of personalization driven by learning progress provides conclusive results in terms of both pedagogical effectiveness (progress observed pre- and post-intervention) and efficiency (learning experience and motivation elicited post-intervention).
Indeed, LP-based personalization (ZPDES driven) provides better learning outcomes and a better learning experience than a linear-path sequence. Furthermore, we observe a synergic effect between LP-based curricula and the ability to express choice (allowing to express self-determination), in accordance with the LP-model. On the contrary, allowing choices, as a form of gamification, can have a deleterious effect and act as a distractor when combined with linear-path curricula. 

Other results in field studies have shown the LP approach to be effective for students with specific learning needs (i.e., Autism and intellectual deficiency,\cite{mazon2022pilot} ) and applicable to different domains (health education, \cite{delmas2018fostering}) showing the generality and promising perspectives of the approach. Future work could also evaluate the use of this approach in the field of cognitive training to improve the number of responders to training in various samples (age, neurodiversity).

%% file: methods.tex
\section{Materials and Methods}

\input{algo}

\input{scenario}

\input{experimentals}

%% file: algo.tex

\subsection{Algorithmic definitions}
\label{sec:algDef}

Several systematic reviews or studies on ITS\cite{SRhew2013use, SRgerard2015automated, SRfaber2017effects, SRbartolome2018personalisation, SRiterbeke2021effects} efficacy pinpointed methodological limitations of this new empirical field (no control group, no initial group equivalence, no pre- and post-intervention measurements, etc\cite{SRCHEUNG201388}) and the great variability of the ITS designs or of their use making it difficult to identify which of the ITS features and/or which of the conditions of learning context of their use \footnote{for instance, ITS can be used alone or mixed with a specific teacher-based instructional setting} are critical for successful personalized learning. So in this section, we describe both the ITS features as well as the experimental protocol. As the ITS system presented here was conceived in the context of a project called 'KidLearn', we also refer to this ITS as the KidLearn system. 

The purpose of an ITS is to enable a learner to acquire knowledge and skills related to a specific domain. Modelling such domain is a difficult problem that has been the subject of numerous research\cite{SRclancey1985acquiring, SRlevesque1986knowledge, SRrussell2009upper}. Difficulties also arised during the research around the use of Q-matrix for the use of Multi-Armed Bandit (MAB) for ITS\cite{SRclement2015}. Even if a lot of research has been done to create tools such as Cognitive Tutor Authoring Tools (CTAT)\cite{SRaleven2013knowledge} to help experts create Q-matrix\cite{SRsottilare2016design}, their use in the conception of the domain model can lead to practical difficulties such as human errors, misspecifications\cite{SRnajera2021balancing} and heavy time consumption for the pedagogical expert.  The following Activity Space formalism is defined to address these issues.

\subsubsection{Activity Space}
\label{sec:ActSpace}

A pedagogical Activity Space is considered to be a set of activities that a learner can practice to acquire skills or knowledge components. An activity or exercise is characterized by multiple parameters $a_i$ (difficulty, shape, type, ...) which can take different values $v_{j}$. For example, to work on mathematical skills, an exercise may have a type that works the addition and another type that works subtraction. ``Addition'' and ``subtraction'' are then two possible values for the parameter ``type of exercise''. 

These parameters and their respective values define all the possible activities that can be instantiated inside the activity space. Depending on their nature and meaning, these parameters can be organized in different groups. Such group of parameters is noted as $H_x = {a_{1},\ldots,a_{n_{x}}}$. In addition, these parameter groups can be structured hierarchically, since some parameters depend on others to be used in an activity.
 
Different types of exercises can require different skills, so the first group of pedagogical parameters (which will be the first level in the hierarchy) determines which type of exercise is selected. Several difficulty levels exist for each type of exercise, so different groups of parameters will determine which difficulty is chosen depending on the type of exercise (second level in the hierarchy). In this case, when one exercise type is selected, the parameter groups that determine the difficulty for the other types are not involved in the parametrization of the activity. Thus, not all parameter groups are necessarily used to define all the activities in the activity space. Therefore, an Activity Space is defined as a set of $n_H$ hierarchical groups of parameters, $A = {H_1,\ldots,H_{n_H}}$.

An activity/exercise $e$ is characterized as a particular combination of parameter values inside an activity space where values were selected for each hierarchical group of parameters involved. All the parameters needed to define an activity are instantiated to produce a unique combination of parameter values. The index $u_i$ corresponds a selected value $v_{u_i}$, for a parameter $a_i$, used to generate an activity. To simplify the notation, $u_i$ is noted as a given parameter selected value to generate an exercise to differentiate it with $v_j$ which defines any values of a parameter. 
For a group $H_x$ with $m$ parameters, the selection of each parameter value produce a combination leading to a singular instantiation of this group $h_x = {u_{1}, \dots , u_{m}}$.

\begin{figure*}[ht]
	\centering
		\includegraphics[width=0.5\columnwidth]{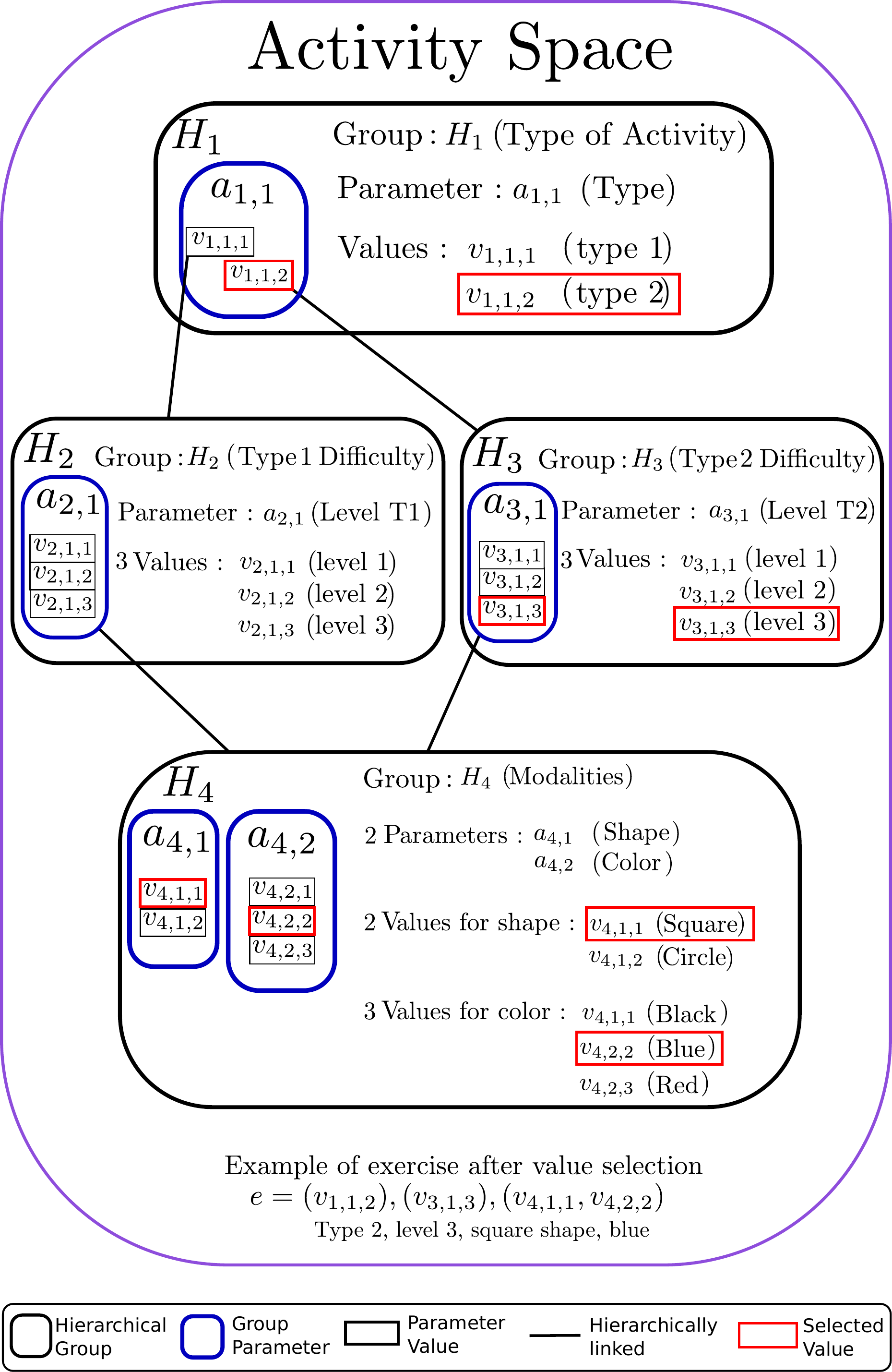}
        \qquad
        \begin{minipage}{0.42\linewidth}\vskip-350pt  To compute an exercise, the values are first selected in the primary group $H_1$. Here 2 values are possible for 1 parameter. 
        $v_{1,1,2}$ is selected thus $u_{1,1} = 2$ and the first group instantiation gives $h_1 = v_{1,1,2}$, meaning the exercise type is 2. This value is hierarchically linked to the group $H_3$. ($H_2$ is not used in this case).\\
        
        Then, values in the group $H_3$ are selected. Here, $v_{3,1,3}$ is selected so $h_3 = v_{3,1,3}$, i.e. a difficulty level of 3. All levels are linked to the group of modalities $H_4$.
        Next, values are selected in the group $H_4$. For example $v_{4,1,1}$ and $v_{4,2,2}$ leading to $h_4 = v_{4,1,1}, v_{4,2,2}$ meaning a square shape and a blue colour.\\
        
        So the final exercise resulting from the values selection is: 
        
        $e = h_1, h_3, h_4 \leftrightarrow e = (v_{1,1,2}), (v_{3,1,3}), (v_{4,1,1}, v_{4,2,2})$\\ 
        
        This means the exercise is of Type 2, level 3 with square shape and blue modalities.
        The Activity Space groups all the activities that can be generated in this way.
        \end{minipage}
        \caption{Illustration of an Activity Space with 4 groups of parameters, and a selection of values which lead to an example of an activity. A group is noted $H_x$, a parameter $a_{x,i}$ and a value $v_{x,i,j}$.}
	\label{fig:ActSpace}
\end{figure*}

After the selection process, a certain number of groups was instantiated, each producing an activity $e = {h_1, \dots , h_{n_e}}$, which groups all parameter values that were selected to produce a unique combination. An activity space groups all possible distinct combinations of parameter values that can define an activity in this space. An illustration of a simple example of an Activity Space with the instantiation of an exercise is shown in figure~\ref{fig:ActSpace}. 

How can this activity space be managed to propose relevant and personalized activities and offer a motivating and enriching experience to the learners ? Several methods are proposed below to answer this question.

\subsubsection{ZPDES : a combinaison of Multi-Armed Bandit and Intrinsic Motivation theories to manage Teaching Sequences}
\label{sec:ZPDES}

To address the challenge of managing activities in an Intelligent Tutoring System, the ZPDES (Zone of Proximal Development and Empirical Success) has been proposed \cite{SRclement:tel-01968241, SRclement2015}.
It relies on state-of-the-art Multi-Armed Bandit techniques (MAB) \cite{SRauer2003nonstochastic,SRbubeck2012MAB} and exploit the empirical estimation of learning progress\cite{SRoudeyer2007intrinsic} to manage the Activity Space.

To use a casino analogy, multi-armed bandits describe the problem of finding the slot machine that provides the maximum reward, initially unknown, in a set of many different  machines. To find the best machine it is needed to spend money exploring each one before being able to always bet on the best one. This boils down to what is called the ``exploration/exploitation'' trade-off in machine learning and  learning processes generally.
Here, these approaches are adapted to ITS where the gambler is replaced by the activity manager, the choice of machine is replaced by a choice of activity parameter values, and the reward is replaced by the student learning progress.
It is assumed that activities which are currently estimated to provide a good learning progress must be selected more often. Prior work showed that this assumption holds for many classes of problems \cite{SRLopes12ssp} and is intrinsically motivating for people \cite{SRgottlieb2013information}. 

A particularity here is the reward (learning progress) which is non-stationary. This requires specific mechanisms to track its evolution. Indeed, a given activity will stop providing a reward, or learning progress, after the student reaches a certain mastery level of the skill or of the activity. Also, it cannot be assumed that the rewards are independent and identically distributed as different students will have different preferences, sensibilities or human factors. They may be distracted or make mistakes when using the system which can create spurious effects. Thus, the framework introduced here rely on a variant of the EXP4 algorithm, proposed initially by \cite{SRauer2003nonstochastic}, which considers a set of experts\footnote{The general term ``expert'' \cite{SRcesa1997use} is used to refer to strategies used in algorithms for``prediction with expert advice'',  ``by combining the predictions of several prediction strategies''.} 
and make a choice based on the proposals of each expert. In case presented here, the experts are a set of variables that track how much reward each activity is providing \cite{SRLopes12ssp}. These bandit experts are used to evaluate the quality of each activity parameter value during the learner's working session. 

A set of simultaneous MAB is used for each group of parameters in the activity space rather than a unique MAB used for each possible combination due to the combinatorial explosion of parameter values. The first alternative of considering a given arm for each activity would increase the number of arms. That would increase the number of parameters and the number of trials required to estimate learning progress and thus the learning time. Also, the approach presented here allows the algorithm to identify which features benefit some students more than others.

For example, to learn a particular skill, the same information may be presented in a written text, a video, a game, an audio track or another format. The knowledge the learner must acquire is the same in each case, but the format of the information differs and individual learners may be more receptive to a particular format.

A case can be imagined where a student works to learn mathematics; different activities are presented to him in a written format, and he almost never answers correctly. But when activities are presented to him in an audio format, he begins to succeed and progress. In this case, the problem is not about the mathematics skills he could learn, but rather his skills in reading. As another example, if an audio format is presented to a student with a hearing impairment, he will not perform and progress as well as with a written format. In light of this, the introduced method evaluates the relevance of and gives meaning to each feature and detects weaknesses and preferences of each student. The propositions it makes are more customized than the ones from an approach where particular combinations would be evaluated, but where features are not taken into account.

Each simultaneous MAB, used to sample each group of parameter, uses a bandit algorithm derived from EXP4\cite{SRLopes12ssp}. The following process is described in Alg.~\ref{alg:sampleValue}. 
\begin{algorithm}
\caption{Procedure to stochastically sample group parameter values according to their quality evaluation.}
\begin{algorithmic}[1]
\Require Group $H_x$ of $m$ parameters $a_i$ with their $n_i$ values $v_{j}$
\Require Set $W_{x}$ of $m$ experts $w_{i}$ for each parameter
\Require $\gamma$ rate of exploration
\Require distribution for parameter exploration $\xi_u$
\Procedure {sampleValues}{$H_x$, $W_{x}$}
\For{$i=1\ldots m$}
      \State $\tilde{w_{i}} \gets \frac{w_{i}}{\sum^{n_{i}}_{j=0} w_{i}(v_{j})}$
      \State $p_i \gets \tilde{w_i} (1-\gamma)+ \gamma \xi_u$ (Eq.~\ref{eq:probExpert})
      \State $u_i \gets$ value sampled from $a_i$ proportionally to $p_i$ 
\EndFor
\State $h_x \gets \{u_1,\dots, u_{n_x}\}$
\State \textbf{return}  $h_x$
\EndProcedure
\end{algorithmic}
\label{alg:sampleValue}
\end{algorithm}

For each parameter $a_i$ inside a group, the quality of its values is evaluated by a bandit expert $w_i$. An expert track the reward provided by each value $v_j$ on the last several sampling to compute its quality noted $w_{i}(v_{j})$.

At any given time, the value to use for each parameter is sampled according to the probabilities given by: 
\begin{equation}
	p_{i} = \tilde{w_{i}} (1-\gamma)+ \gamma \xi_u
\label{eq:probExpert}
\end{equation}
where $\tilde{w_{i}}$ are the normalized $w_{i}$ values to ensure a correct probability distribution, $\xi_u$ is a uniform distribution that ensures sufficient parameter exploration and $\gamma$ is the exploration rate, tuned to make the exploration wide or narrow. This sampling methodology leads to stochastically select a value, proportionally based on its quality and $\gamma$. For low values of $\gamma$, the parameter value is chosen mostly based on its quality, whereas for high values of $\gamma$, low quality parameter values have a higher probability of being picked, which means a high exploration rate.
The set of experts correlated to $H_x$ is noted $W_x = w_{1}, \dots, w_{n_x}$ . From now on, a Stochastic Activity Space $A^S$ is considered to be a set of tuples $(H_x, W_x)$. 

\begin{figure*}[ht]
  \centering

  \includegraphics[width=0.5\columnwidth]{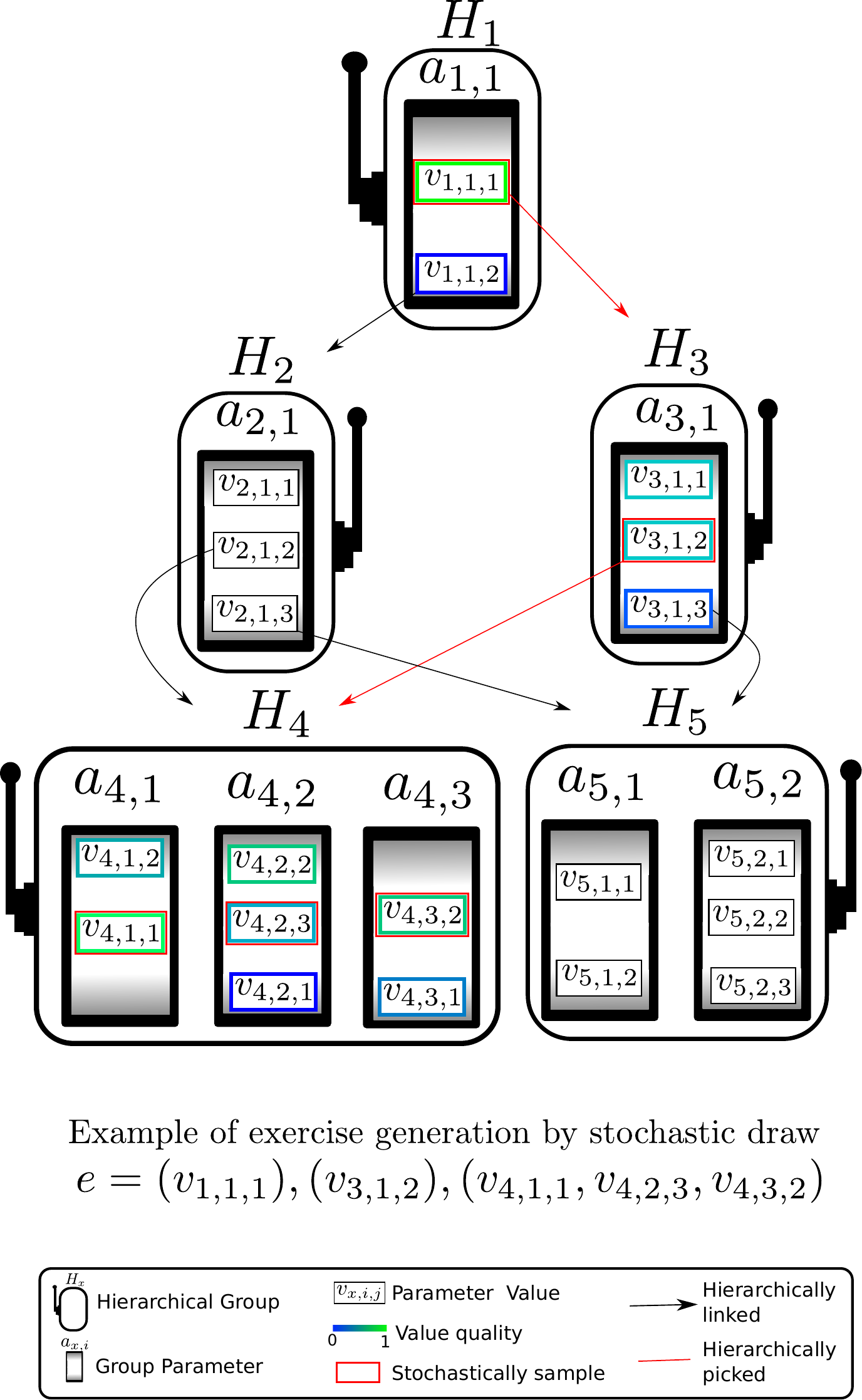}
        \qquad
        \begin{minipage}{0.42\linewidth}\vskip-350pt
        The primary group $H_1$ has one parameter $a_{1,1}$, for this parameter, the first values is evaluated to have a higher quality than the second one: $w_{1,1,1} \ge w_{1,1,2}$. There are then more chances for the first value to be sampled. Here, the result of the stochastic sample is that $v_{1,1,1}$ is drawn. 

$v_{1,1,1}$ is linked hierarchically to $H_3$, which is the next to be instantiated. The two first values have the same medium quality (same chance to be drawn), and the last one has a low quality (less chance of being drawn). $v_{3,1,2}$ is drawn, which has a dependency with the group H4.\\

H4 is then instantiated and has three parameters. The first parameter $a_{4,1}$ has its first value evaluated to be more interesting than its second one. The quality for the second parameter is ordered as $w_{4,2,2} \ge w_{4,2,3} \ge w_{4,2,1}$ and for the third parameter $w_{4,3,2} \ge w_{4,3,1}$. \\

The three parameters are sampled simultaneously, and $v_{4,1,1}$, $v_{4,2,3}$ and $v_{4,3,2}$ are drawn. Even though $w_{4,2,2} \ge w_{4,2,3}$, $v_{4,2,3}$ had a chance to be drawn, following the process of exploration. The result of the activity generation is:

\noindent $e = (v_{1,1,1}), (v_{3,1,2}), (v_{4,1,1}, v_{4,2,3}, v_{4,3,2})$. 
\end{minipage}

\caption{Hierarchical Multi-Armed Bandit with 5 groups of parameters and a selection, by stochastic draw, of an example of activity. A group is noted $H_x$, a parameter $a_{x,i}$ and a value $v_{x,i,j}$.}
  
  \label{fig:HMAB}
\end{figure*}

To generate an activity, this process is done recursively on the hierarchical groups that are involved in the activity generation, in accordance with the hierarchical dependencies between the groups of parameters. As describe in Alg.~\ref{alg:genActivity}, it starts by the instantiation of the primary group of parameter $H_1$ and is followed by the instantiation of the groups that are iteratively selected according to their dependencies. This leads to a stochastic draw of activity, resulting from the combination of each parameter value sampled depending on the evaluation of their quality by each expert. An abstract illustration of an activity generation is presented in figure~\ref{fig:HMAB}. 
\begin{algorithm}
\caption{Activity generation procedure based on an Activity Space and Hierarchical Multi Armed-Bandit mechanisms.}
\begin{algorithmic}[1]
\Require A Stochastic Activity Space $A^S$, set of tuples $(H_x, W_x)$
\Procedure {genActivity}{$A^S$}
  \State \{Initialize\} 
  \State $i \gets 1$
  \State Instantiate primary group $h_1$:
  \State \hskip\algorithmicindent $h_1 \gets sampleValues(H_1, W_{1})$
  \State \{Recursive sample\} 
  \While{$h_i$ require to instantiate a group $H_x$}
  	\State $i \gets x$ 
      \State $h_i \gets sampleValues(H_{x}, W_x)$  
  \EndWhile
\EndProcedure
\State \textbf{return}  $e = h_1, ..., h_i$
\end{algorithmic}
\label{alg:genActivity}
\end{algorithm}

Once an activity is generated\label{sec:rewardDef}, this activity is proposed to a learner to work on and answer to. After answering, the algorithm retrieves his answer. Each time an exercise is given and answered, the expert of each parameter value $u_{i}$ used in the activity is updated:

\begin{equation}
  w_{i}(u_{i}) \gets \beta w_{i}(u_{i}) + \eta r
  \label{eq:expertUpdate}
\end{equation}
where $r$ is a reward that measures the benefit the activity gives to the learner in terms of progress. The variables $\beta$ and $\eta$ define the tracking dynamics of this estimation, which is the compromise between the old rewards and the new ones brought by the last activity. This mechanism allows the experts to assess and update the quality of each parameter value, used over time, based on the student learning. 

As discussed before, focusing on activities that are providing more learning progress can act as a strong motivational cue \cite{SRgottlieb2013information}. Equation~\ref{eq:calculRewZPDES} describes the reward computation which is based on estimating how the success rate on each parameter group is improving :
\begin{equation}  
    r_x = \sum_{t=T-d/2}^t \frac{C_t}{d/2} - \sum_{t=T-d}^{T-d/2} \frac{C_t}{d -d/2} 
    \label{eq:calculRewZPDES}
\end{equation}
where $C_t=1$ if the activity at time $t$ was solved correctly. At the time $T$, the equation compares the success of the last $d/2$ samples with the $d/2$ previous samples, providing an empirical measure of the time evolution of the success rate. 

This reward allows to compute a measure of the quality of each activity parameter value, measuring how much progress it provided in a recent time window. Both extreme cases, when an activity is already mastered or when it is impossible to solve, will have a reward of zero. Moreover, parameter values providing a faster progress are assumed to be better than others. The algorithm to compute the reward is presented in Alg.~\ref{alg:computeRewZPDES}.

\vspace{0.1cm}

\begin{algorithm}
\caption{ZPDES reward computing procedure}
\begin{algorithmic}[1]
\Require Activity $e$
\Require Student answer $C$
\Require parameter $d$
\Procedure {computeReward}{$e$, $C$}
    \For{$h_x$ in $e$}  
        \State $r_x =  \sum_{t=T-d/2}^t \frac{C_t}{d/2} - \sum_{t=T-d}^{T-d/2} \frac{C_t}{d -d/2}$ (Eq.~\ref{eq:calculRewZPDES})
    \EndFor
    \State $r \leftarrow {r_1, \dots, r_{n_e}}$
    \State \textbf{return} $r$
\EndProcedure
\end{algorithmic}
\label{alg:computeRewZPDES}
\end{algorithm}

A pure selection, based solely on the previous considerations, would explore all possible activities that could be generated in the activity space from the start of the work process. This would have two drawbacks. First, the type and difficulty of the exercises proposed could change too often and reduce the learners' motivation and engagement. 
Second, it might not be possible to explore all activity parameters to estimate the learning progress they are providing. To ensure that learners remain in challenging but possible to achieve areas and to be able to assess the quality of each parameter, a mechanism to limit exploration is introduced. 

\begin{figure}
\centering
  \includegraphics[width=0.4\linewidth]{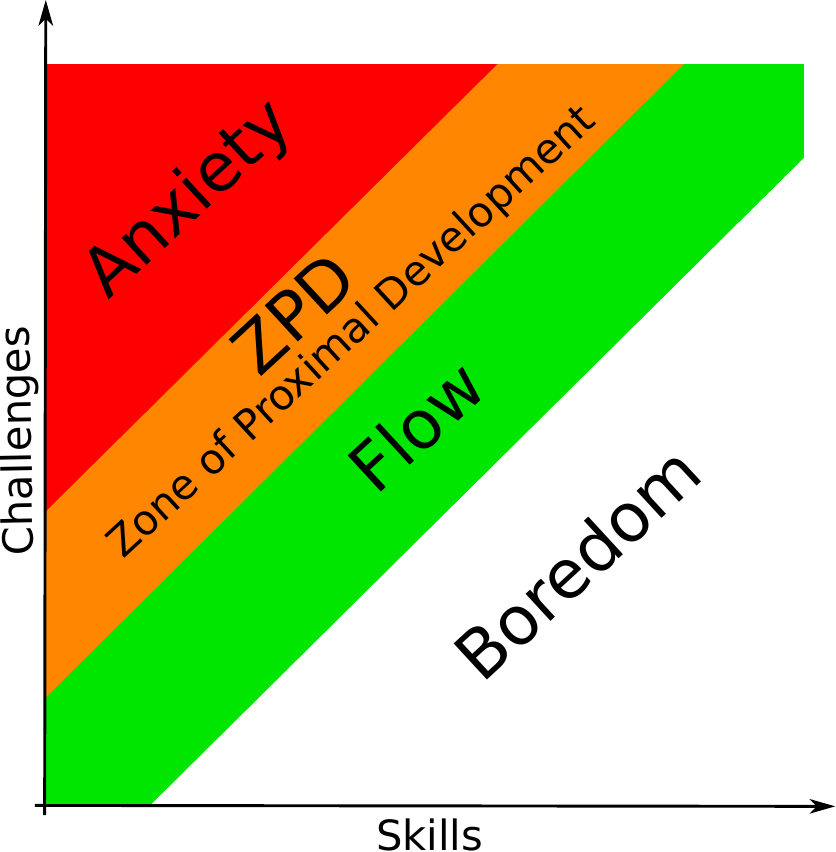}
  \caption{The concept of Zones of Proximal Flow\cite{SRbasawapatna2013zones} presents the idea of the Zone of Proximal Development being located in between regions of Flow and anxiety.}
  \label{fig:ZPF} 
\end{figure}

Inspired by the Zone of Proximal Development theory \cite{SRvygotsky1978mind} and the concept of Flow \cite{SRcsikszentmihalyi1975beyond}, a pedagogical expert has the possibility to specify rules that define an evolving set of possible/activated activities, judged relevant for the student. These activities keep the student in the zone of Flow or in the Zone of Proximal Develpment (ZPD) based on his successive results (see Fig.~\ref{fig:ZPF}).
The goal is to propose activities that are neither too easy nor too difficult, without having to try all possible activities. The different possible activities proposed by the algorithm are then the active ones which are inside the ZPD. The use of the ZPD offers three advantages: it helps to improve motivation as discussed before, it further reduces the need of quantitative metrics for the educational design expert and it provides a more predictive choice of activities.

The implementation of these principles\label{sec:zpdDef} is applied to the algorithm by the definition of rules that guide the bandits experts and restrict the exploration of the activity parameters. Theses rules define activation/deactivation mechanisms which allow the algorithm to activate and deactivate parameters values, depending on the evaluation of their relevance and the quality of the students learning process. As a consequence, the active parameters values generate a subset of all possible activities inside the activity space. The ZPD is defined here as a particular subset of active activities with its corresponding parameter values. 

Following this principle, ordered relations between activity parameter values can be defined, leading to a ``graph'' governing the activity space which are combined with the set of rules that define and manage the ZPD using activation/deactivation mechanisms. 
Rules and ordered relations are not always defined for each parameter: there is a distinction between subsets of activity parameters that have a clear difficulty progression, and subsets that don't. For the example used from \ref{sec:ActSpace}, the difficulty levels have a clear ordering while the modalities don't. In practice, the management of the ZPD proceeds as follows. For activity parameters with no difficulty level relations between their values, a free exploration is allowed and so all of their values are always active. While for parameters that have a clear progression in difficulty, the values will be activated and deactivated depending on the success rate over all active values. 
The following mechanism is proposed to generally manage the ZPD. When the recent learner success rate over all active parameter values $\delta_{i,ZPD}$ reaches a value $\lambda_{ZPD}$, the ZPD is expanded to explore another parameter value $v_{i,j}$ by initializing its expert as : $w_i(v_{i,j}) = \min w_i(v^{ZPD})$. 

When the recent success rate for a particular value $\delta_{v_{i,j}}$ is higher than a threshold $\lambda_d$, this activity can be deactivated and removed from the active list of values. These two threshold allow to configure the general exploration behaviour of the algorithm inside the activity space and is illustrated in figure~\ref{fig:ZPDESexplore}.

The main intuition of this process is that when there are some activities whose difficulty grows, the ZPD will have to grow at the same rate. When activities do not have a clear order of difficulty, or when the order might change from person to person, then it is necessary to allow wider exploration of the activities to accommodate individual differences.
Another kind of mechanism is added to allow a more precise and specific parametrization of the ZPD. Indeed, the algorithm needs to be able to activate and deactivate values when the conditions of exploration for an activity parameter depends on another set of parameters. If the value $v_{g,i,j}$ of parameter $a_i$ of group $H_g$ requires a certain mastery level of value $v_{x,y,z}$, a threshold $\lambda_{v_{x,y,z}}$ is defined corresponding to the success rate a learner must reach with activities using $v_{x,y,z}$ to activate $v_{g,i,j}$. The requirements can be multiple, meaning a values activation can depend on multiple other values, parameters or group to be activated.

For example, the difficulty level for a particular type of exercise can require only the previous level to be mastered, or it can require various previous levels, or it can even require other types of exercises in different levels to be mastered. In a mathematics analogy, if a student works on a simple subtraction activity, he needs to master simple addition activities to be able to succeed. And if he works on hard subtractions with basic decimal number, he needs to master hard addition and basic decimal numbers first to succeed. 

But this mechanism can lead to blockages in the exploration. If a type A of exercise is easy during $3$ levels for a student, leading to a $100\%$ success rate, the quality of this type will be very low. The algorithm will then select other types of excercises more often. But if the level $2$ of type B needs a higher level of type A to be mastered, the algorithm will continue to propose more type B without being able to activate the level $2$ until the required level type A is mastered. 

To address this issue, a quality upgrading mechanism is added for values that are required. If ZPDES tries to activate a value parameter but is unable to do so due to a requirement, the qualities of required parameter values are increased. This way, ZPDES will exploit values needed to expanded the graph as a priority.   

Basically, the first mechanism introduced to manage the ZPD is a simplification of the mechanism presented above. It is integrated to reduce the information needed to define ZPD rules and to allow a freer exploration of the activity space by the algorithm. 

\begin{algorithm}
\caption[ZPDES algorithm]{ZPDES algorithm. It manages pedagogical curricula based on an ActivitySpace, a multi-armed bandit algorithm (genActivity procedure), and a set of rules to extend and/or shrink the ZPD where the student evolves.}
\label{alg:SSBanditZPDES}
\begin{algorithmic}[1]
\Require A Stochastic Activity Space $A^S$
\Require $R^{ZPD}$ rules  
\State Initialize bandit experts uniformly according to $R^{ZPD}$.
\While{\textit{learning}}
\State Generate activity $e \leftarrow genActivity(A^S)$ (Alg. \ref{alg:genActivity})
\State Get learner answer $C$
\State Compute reward $r$ $\leftarrow computeReward(e,C)$ (Alg. \ref{alg:computeRewZPDES})
\State {Update greedy expert}
\For{($h_x, r_x$) in ($e, r$)}
    \For{$u_{i}$ in $h_x$}
        \State $w_{i}(u_{i}) \leftarrow \beta w_{i}(u_{i}) + \eta r_x$
    \EndFor
    \State Update ZPD:  
    \State \hskip\algorithmicindent activate/deactivate {$w_{i}$} based on $R^{ZPD}$
\EndFor
\EndWhile
\end{algorithmic}
\end{algorithm}

The final ZPDES algorithm is presented in Alg.~\ref{alg:SSBanditZPDES}. One of the main advantages of these principles is the consideration of an empirical estimation of the learning progress. It has been proposed in artificial curiosity and intrinsic motivation systems\cite{SRoudeyer2007intrinsic}. Instead of relying on a precise model of the learning system, with all limitations in terms of parameter identification and computational complexity, it is possible to create surrogate functions of the learning progress. These estimators are simple, robust, and, even if not optimal, more flexible and adapt better to model errors and situations where the model assumptions are violated.

\paragraph{Added value of the LP based approach}  

Presenting the best activities to a learner at a given time to stimulate his learning as well as motivation is a crucial issue in ITS design. Therefore, the evaluation of the student knowledge level, i.e the "student model" (and subsequent adaptations), needs to be accurate over time to provide the best match between the learning activity and the learner's zone of proximal learning (\cite{SRmetcalfe2020epistemic}). 

The general principle of the LP-based approach is to propose to each learner the activities that maximize his or her progress within the ITS activities. Such an adaptation is dynamic and depends on the learner's performance. 

The activities are structured as a graph into an activity space based on expert knowledge (i.e, the "domain model"). The learning paths are then personalized in two ways; first, by exploring and testing continuously various activities inside the activity space in order to assess their didactic potential for the learner's progress in real time; second, by exploiting and mainly proposing the activities identified as being the most effective for him/her based on the previous assessment.    
The simplicity of the activity space on which the approach is based is a first asset. Indeed, it does not rely on any multidimensional student or domain models, and then only the learner's information about the estimated learning progression for each activity is required. 

A second asset of LP approach is to leverage an efficient and simple optimization method consisting of prioritize activities' parameters identified as yielding significant learning outcomes, i.e. Zone of Promixal Development (ZPD). For that, thanks to our multi-armed bandit algorithm, at each step of the optimization process, one arm is chosen and the resulting payoff is estimated, the objective being to discover dynamically the best arm for each student. 
So, various and heterogeneous paths are possible across students as expected for taking into account  specific learner's need at a specific point in time. So, the LP approach empowers the ITS adaptivity for diagnosing learner's ZPD and for making appropriate adjustments to the specific learners' needs.
Taken together, these two main assets enable to quickly design successful ITS (albeit substantial efforts must be done to parameterize the activity graph on which ZPDES runs) and are supportive to hybrid system combining machine learning and rule-based approaches \cite{SRfournier2010its}.

\subsubsection{Predefined Sequence}
\label{sec:PredefSeqSec}

To be able to evaluate ZPDES, a baseline has been built as the form of a Predefined Sequence (Predef). This Predefined Sequence is a simple algorithm inspired by mastery learning strategy \cite{SRbloom1968learning} and based on instructional design whose reliability has been validated through several user studies \cite{SRroy12math}. It does not use any machine learning technology.
It consists of a sequences of predefined activities organised by difficulty. When a student is working an activity, he needs to have 3 success out of 4 exercises to pass to the next activity, or else he stays on the same activity. The sequence of activities used in this experiment as been designed by an expert in teaching of mathematics for primary school. The sequence is composed of 27 activities we can divide into 8 groups. Each group corresponds to a type with or without decimal. (M, R, MM and RM with integers only, then M, R, MM and RM with decimals). 
Table~\ref{tab:PredefinedSequence} shows the $27$ successive activities for the students following the parameters defined in section~\ref{sec:kidlearn}. 
\begin{table*}[h!]
  \centering
  \small
    \begin{tabular}{l|c|c|c||c|c|c|c||c|c||c|c|c|c}
                & G1.1& G1.2& G1.3& G2.1 & G2.2 & G2.3  & G2.4  & G3.1& G3.2& G4.1& G4.2& G4.3& G4.4  \\\hline
      Ex Type   & M   & M   & M   & MM   & MM  & MM     & MM    & R    & R  & RM  & RM  & RM  & RM    \\\hline
      Difficulty& 1   & 2   & 3   & 1    & 1   & 2      & 2     & 1    & 2  & 1   & 1   & 2   & 2     \\\hline
      Cents Not & -   & -   & -   & -    & -   & -      & -     & -    & -  & -   & -   & -   & -     \\\hline
      Remainder & -   & -   & -   & -    & -   & -      & -     & -    & -  & -   & Int & -   & Int   \\\hline
      Money Type&Real & Real&Real & Real & Real& Real   & Real  & Real &Real&Real &Real & Real& Real  \\\hline
    \end{tabular}
    
    \vspace{0.5cm}

    \begin{tabular}{l|c|c|c|c||c|c|c|c}
                    & G5.1      & G5.2      & G5.3      & G5.4      & G6.1& G6.2& G6.3& G6.4    \\\hline
        Ex Type     & M         & M         & M         & M         & MM  & MM  & MM  & MM      \\\hline
        Difficulty  & 4         & 5         & 5         & 6         & 3   & 3   & 4   & 4       \\\hline
        Cents Not   & x\euro x  & x\euro x  & x,x\euro  & x,x\euro  & -   & -   & -   & -       \\\hline
        Remainder   & -         & -         & -         & -         & -   & Int & -   & Dec     \\\hline
        Money Type  & Real      & Real      & Real      & Real      & Real& Real& Real& Token   \\\hline
    \end{tabular}

    \vspace{0.5cm}
        
    \begin{tabular}{l|c|c|c||c|c|c|c}
                    & G7.1      & G7.2      & G7.3    & G8.1& G8.2& G8.3& G8.4   \\\hline
        Ex Type     & R         & R         & R       & RM  & RM  & RM  & RM     \\\hline
        Difficulty  & 3         & 3         & 4       & 3   & 3   & 4   & 4      \\\hline
        Cents Not   & x\euro x  &x\euro x   & x,x\euro& -   & -   & -   & -      \\\hline
        Remainder   & Int       & -         & Int     & -   & - & Int   & Dec    \\\hline
        Money Type  &Real       & Real      & Real    & Real& Real& Real& Token  \\\hline
\end{tabular}  
    
    \caption{Detailed predefined sequence}
    \label{tab:PredefinedSequence}
\end{table*}

\subsubsection{Choice}
\label{sec:choice}
\begin{figure}[h!]
\centering
  \includegraphics[width=0.6\linewidth]{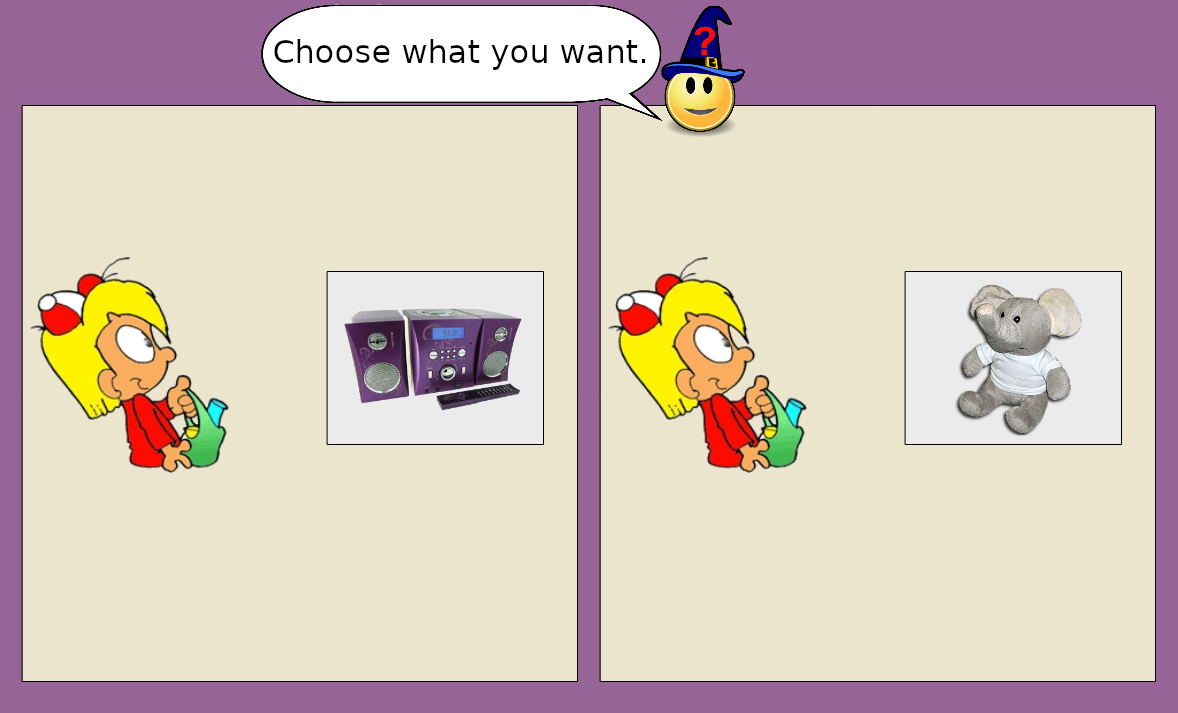} 
  \caption[Object choice interface]{\label{fig:objChoice}Object choice interface. The student choose the object(s) he wants to train with by typing on it. The bubble indicate instruction ``Choose what you want''.)}
\end{figure}

The ability to make one's own decisions, i.e. the ability to make choices, is part of the learning-progress theoretical framework\cite{SRoudeyer2007intrinsic2,SRlopes2012strategic,SRoudeyer2016intrinsic} on which the development of the ZPDES algorithm was scaffolding. Choice expression was also shown to have a positive motivational impact and an efficient vector of performance \cite{SRleotti2011inherent, SRmurayama2013self, SRcordova1996intrinsic}. As ZPDES and Predef do not actually enable students to express choices (which are performed by the machine learning algorithm), two conditions are introduced to reintroduce this ability and study its impact. PCO and ZCO do not change the way ZPDES and Predef control the evolution of parameters of learning activities, but they introduce contextual choice on the objects used to instantiate visually the learning activities the students train as presented in figure~\ref{fig:objChoice}. The aim is to increase intrinsic motivation by adding a preference depending on the student personality and thus introducing an emotional and motivational valence on the object \cite{SRcarstensen2003socioemotional}.

As explained before, studying the impact of choice as a motivational and learning tool, and decoupling it with the actual pedagogical content (e.g. difficulty of exercises) are interesting. ZCO is an experimental condition where the student has choice over a contextual parameter: the objects presented on the screen. In the money game scenario (see below \ref{sec:kidlearn}), students compose sums corresponding to the item prices or the change to be given when a customer purchases an item. The choice given to the student is between two different objects, but the activity parameterization is the same. The activity is still selected by a ZPDES algorithm and the choice has no impact on the ZPDES operation. The interface used to implement this experimental condition is shown in figure~\ref{fig:objChoice}. Only the type of exercise and the objects are presented to simplify the interface and reduce the perturbation of the student to a minimum. The position of the choice icon on the screen is determined randomly to avoid presentation bias.

%% file: scenario.tex

\subsection{Kidlearn activities scenario}
\label{sec:kidlearn}

The teaching scenario used here is about the use of money to teach children how to decompose numbers, typically targeting $7$-$8$ year old students. It corresponds to a set of mathematical skills and learning scenario that are part of the official learning curriculum of French primary schools for children of this age range. This scenario was chosen for its simplicity, while remaining rich enough to offer different learning/teaching trajectories to impact individual students differently. The entire conception of exercises, ranging from their parameterization to the visual interface, was conceived in collaboration with a specialist of didactics of mathematics and participatory design of primary school teachers.

Furthermore, combining number and money manipulation is a way to instantiate abstract knowledge into a practical, useful real-world scenario. This scenario is instantiated in a browser environment.

\begin{figure*}[h!]
    \centering

    \bgroup
   \setlength\tabcolsep{0.2\linewidth}.
   \begin{tabular}{cc}
        Type M & Type R \\
    \end{tabular}   
    \egroup
    
        \includegraphics[width=0.5\linewidth]{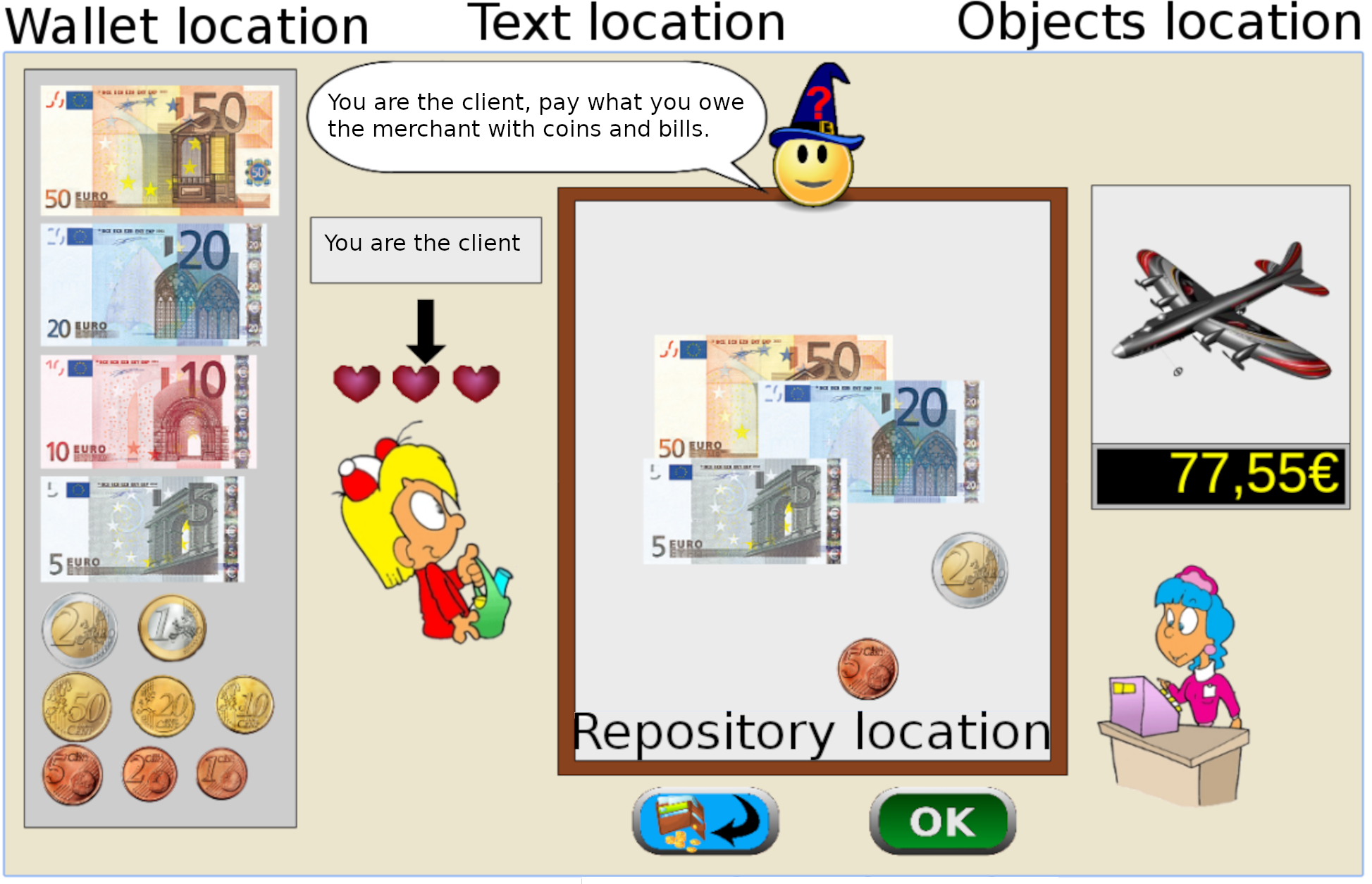}
        {\includegraphics[width=0.5\linewidth]{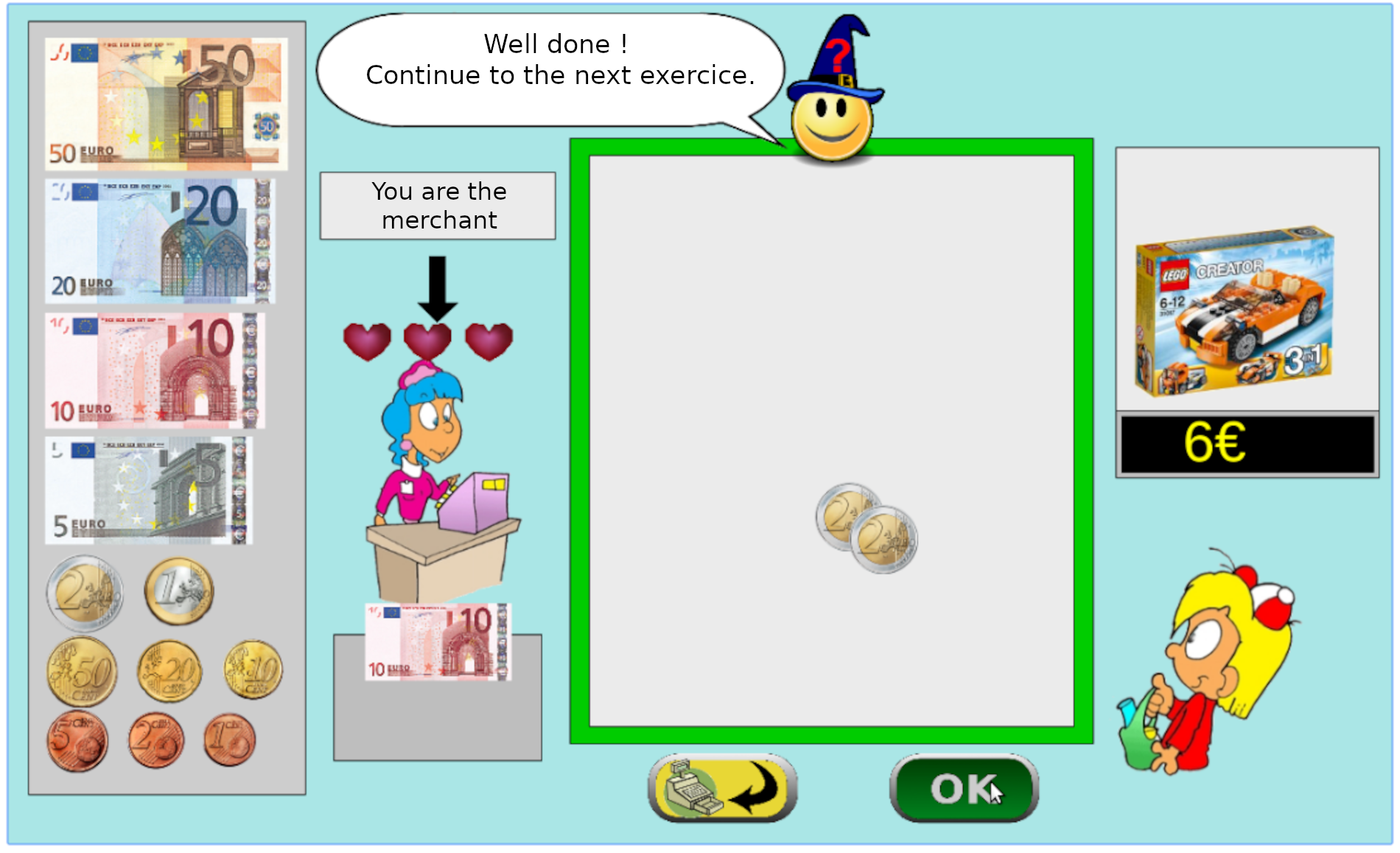}}
        
    \bgroup
    \setlength\tabcolsep{0.2\linewidth}.
    \begin{tabular}{cc}
        Type MM & Type RM \\
    \end{tabular}   
    \egroup   
    
        \includegraphics[width=0.5\linewidth]{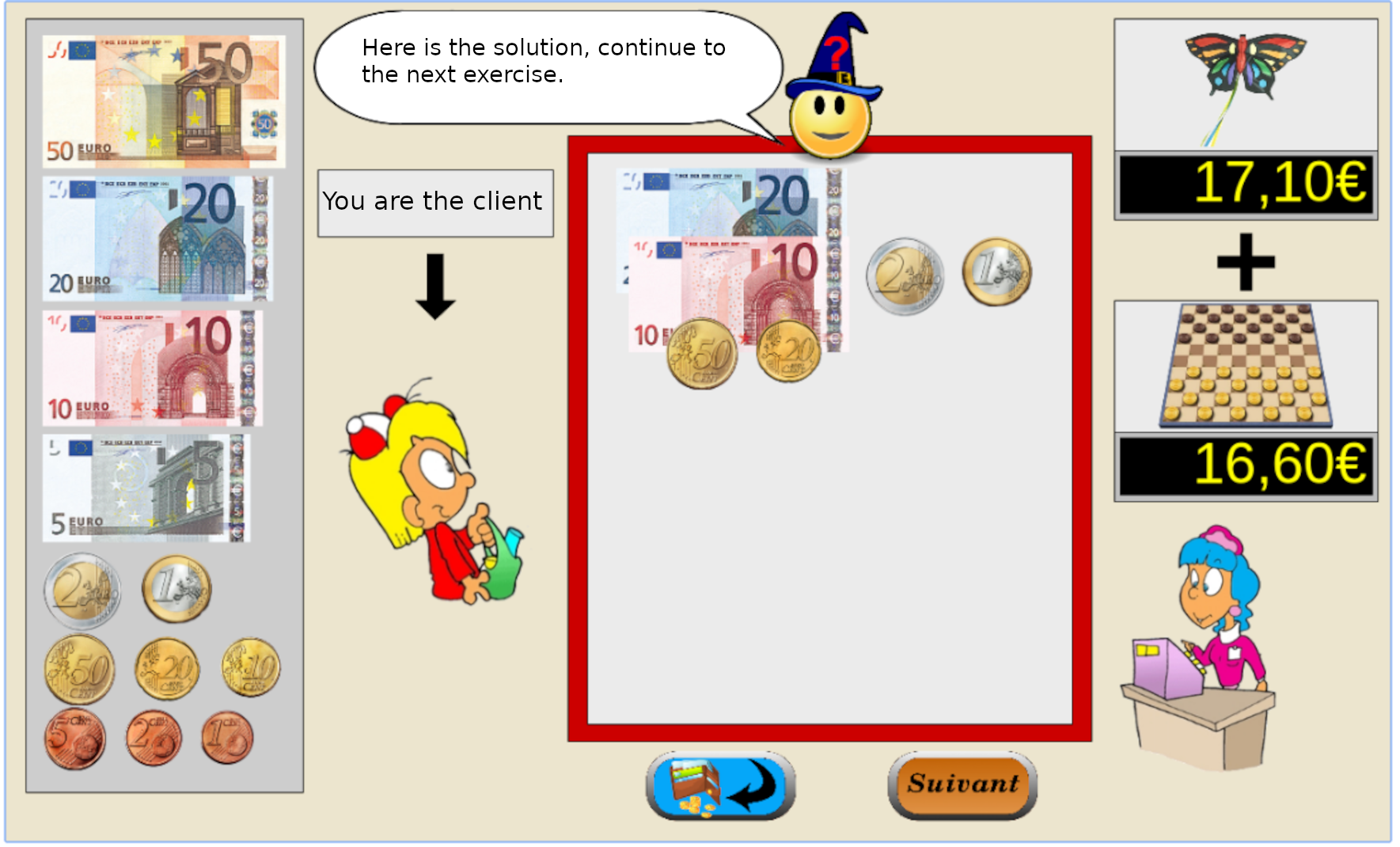}
        \includegraphics[width=0.5\linewidth]{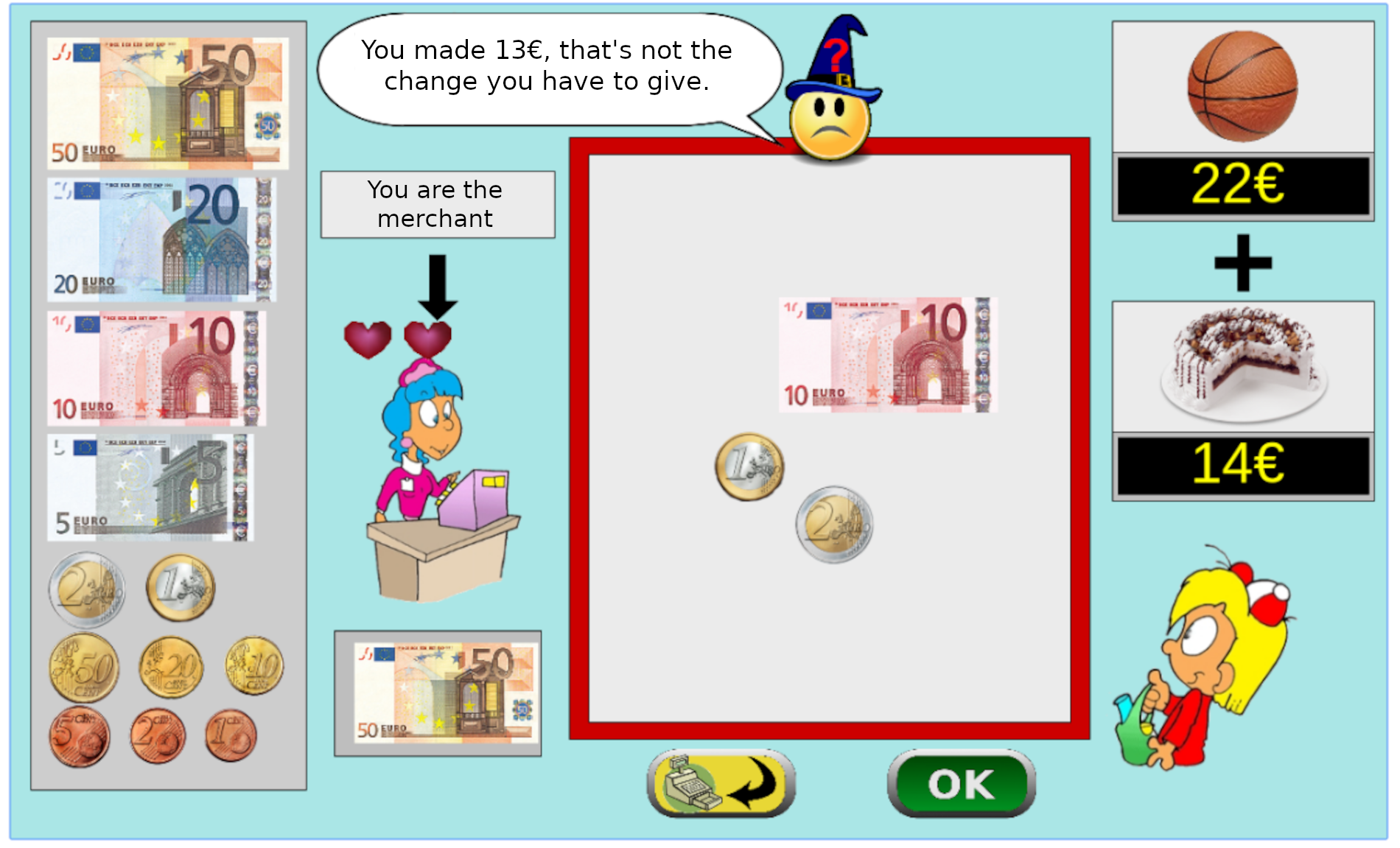}

      \caption{Four principal regions are defined in the graphic interface. The first is the wallet location, where users can pick and drag the money items and drop them on the repository location to compose the correct price. The object and the price are present in the object location. Four different types of exercises exist: M : customer/one object, R : merchant/one object, MM : customer/two objects, RM : merchant/two objects.}
     \label{fig:monnaieinterface}
\end{figure*}

The application proposes exercises to students in the form of money games (see Fig.~\ref{fig:monnaieinterface}). For each exercise type, one object is presented with a given tagged price, and the learner has to choose which combination of bank notes, coins or abstract tokens need to be taken from the wallet to buy the object, with various constraints depending on the exercise parameters. 

\begin{figure*}
    \includegraphics[width=1\linewidth]{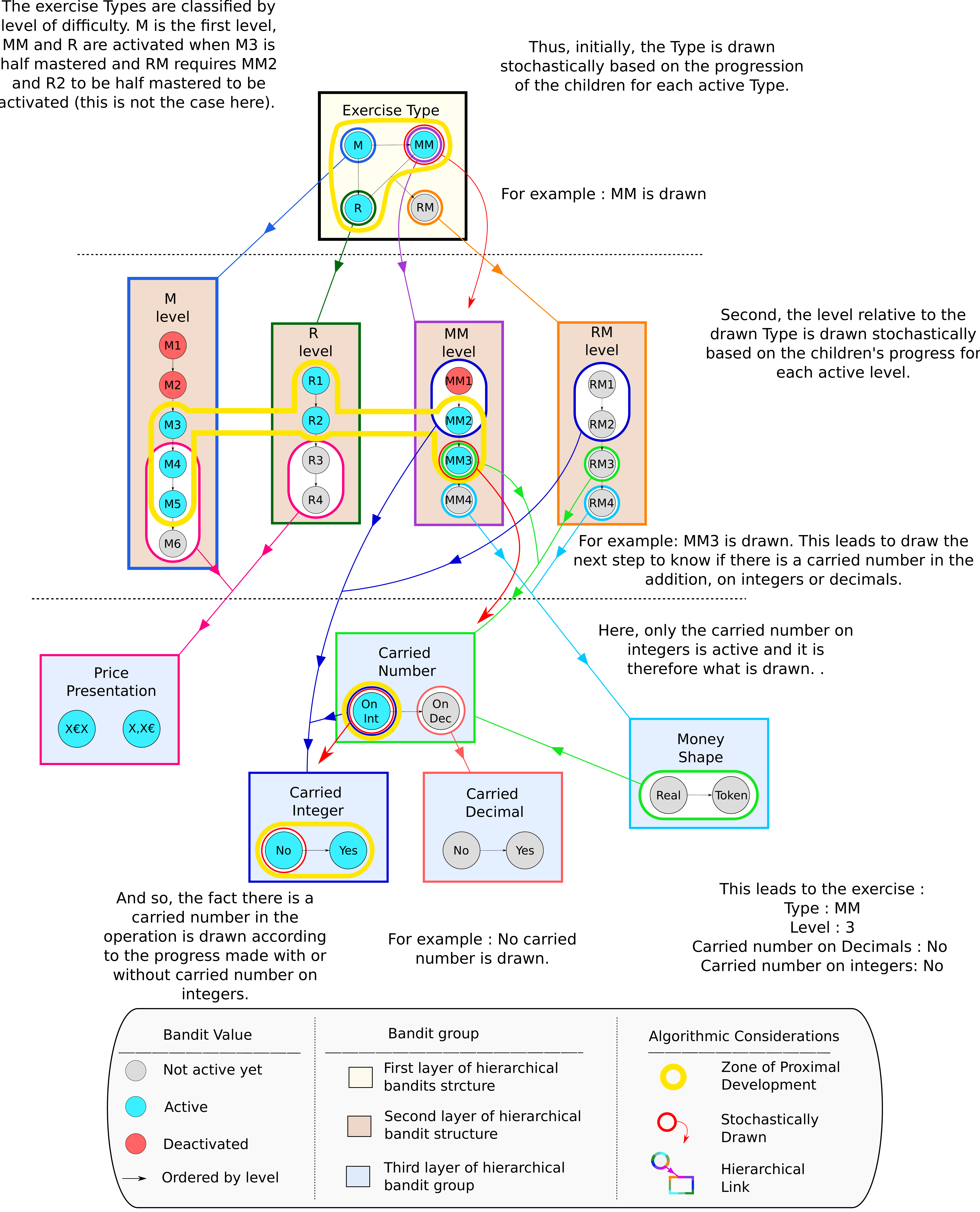}

    \caption{A representation of the activity graph used for a pedagogical scenario to teach child about mathematics by making them manipulate money. An example of activity is sampled at a particular state of the algorithm resulting from activities already made by a hypothetical student.}
    \label{fig:HMABforMoney}

\end{figure*}

The various activities are parameterized using a specific graph summarized in figure \ref{fig:HMABforMoney} with an example of ZPDES activity sampling. 
There are 5 parameters organized hierarchically. First, the \textbf{Exercise Type} is chosen: the student can be the costumer or the merchant and buy or give change with one or two objects, which leads to four different possibilities. 
For each type of exercise, the difficulty is chosen based on the difficulty \textbf{Level} of decomposing a number. A number can be easy to decompose if there is a direct relation with a real bill/coin $a=(1,2,5)$ and hard to decompose if it requires more than one item $b=(3,4,6,7,8,9)$. The exercises will be generated by choosing prices with these properties and picking an object that is priced realistically. A dimension related to the difficulty is the presence of \textbf{Carried Numbers} in the operation, when there are two objects. It is managed by a different parameter because it is not related to a particular exercise type. \textbf{Price Presentation} varies due to the different practices in stores and countries, which do not always follow the standardized rule.  Finally, different \textbf{Money Shapes} are used: Real Euro or poker tokens, which can reduce the visual ambiguity.

Graphical interfaces in ITS can have unwanted side effects. For this reason, the interface was entirely designed with the help of both a specialist of didactic of mathematics and primary school teacher, with several specific design choices motivated by instructional design principles, and motivational and attentional requirements. For example, the interface, shown in figure~\ref{fig:monnaieinterface}, is such that: a) display is as clear and simple as possible; b) there is no chronometer, so that students are not put under time pressure; c) coins and banknotes have realistic visual appearance, and their relative sizes are respected; d) costumer and merchant are represented to indicate clearly the role of the student; e) text quantity is kept to minimum. 

Regarding the interactions with Kidlearn game, the activity starts either with or without a choice for the student between two objects or two groups of objects as in figure~\ref{fig:objChoice}, then one or two objects with their respective price are shown.

To complete the exercise, the student has to drag and drop the money that she/he wants to use from the wallet location to the repository location. 
It is possible to request extra cues, by clicking on the smiley with a hat. They have to click on the ``OK'' button to submit the answer leading to a feedback. If the answer is correct, the feedback is ``Congratulation you can move on to the next exercise''. The experience must provide the most pedagogical gains and so, the student has $3$ opportunities to solve the exercise and extra cues are provided each time the student makes a try. If after $3$ trials the answer is still wrong, then a feedback with the correct solution is given and the system moves on to the next exercise.

The Kidlearn game runs on tablet. The equipment was composed of a set of 30 tablets, two computers and two wifi routers. This equipment was carried in every participating classroom to be independent from school equipment constraints and limit equipment bias.

%% file: experimentals.tex

\subsection{Experimental protocol}
\label{sec:expProtocol}

\subsubsection{Experiment design}
\label{sec:experimentDesign}
According to the Randomized and Controlled Trial (RCT) methods, the experiment was designed to compare four experimental conditions corresponding to two manipulations, i.e. the algorithm conditions, and the object choice and non-choice conditions. A predefined sequence (called "Predef" ) is used as a algorithm baseline with a linear learning path. It is implemented as a series of activities in which the student must have 75\% success over 4 activities of the same type to pass to the next activity type. This Predef sequence was designed by a professional in  didactics of mathematics. The other condition is the ZPDES algorithm condition with personalized paths according to learning progress of student (using a parameterization of learning activities previously designed by the expert).
To estimate the necessary sample size, an a priori power analysis was conducted using G*Power 3.1. 

A recent meta-analysis on studies investigating ITS in learning reported a smallest effect size at g = 0.30 \cite{SRma2014intelligent}. Based on this effect size, with an alpha level of 0.05, the projected total sample size required for detecting a between-between interaction in a repeated measures ANOVA was approximately N = 196 for a power of 0.95, and N = 128 for a power of 0.80.

Within each algorithm condition, we introduced the possibility to self-choose the object of the money exchange for manipulating intrinsic motivation thanks to self-decision making. Hence, four conditions were manipulated as follows: 1) a "Predef" condition with linear path and without self-choice ; 2) a "PCO" condition with linear path and with self-choice of objects; 3) a "ZPDES" condition with Learning-progress based personalizing, but without self-choice of objects  ; and finally 4) a "ZCO" condition with Learning-progress based personalizing with self-choice of objects. 
Thanks to these 4 condition, we were able to assess the effect of Learning-progress based personalizing on learning and motivation performance as well as its possible synergistic effect with playful feature related to the self-choice of object for exercise.

\subsubsection{Participants}
\label{sec:participants}

Teachers from 11 primary schools with 2nd grade classes signed up to participate in the Kidlearn program in Nouvelle-Aquitaine, the South-West region of France. Additionally, we collected the consent from each participant and their parents. 
To assign each student to one of the 4 experimental conditions, randomization was conducted at the classroom level in order to avoid contamination effects.

Although 414 students took the pre-intervention assessment, data from 147 students were excluded from the analysis as they did not complete all the sessions of the experimental protocol (Kidlearn session and/or post-test assessment). In addition, to ensure balance across conditions ex ante, we performed a pseudo-randomized selecting procedure at student level via a computer algorithm. In particular, children were first partitioned into strata according to the following variables: the gender (girl or boy), the child age (7 or 8 years old) and pre-assessment calculation score. Then, within each stratum, children were randomly assigned to a condition. 

The final sample includes 265 children in 24 classes in 11 schools with 62 children for Predef, 59 children for PCO, 76 children for ZPDES and 68 children for ZCO condition. The background characteristics of the students in terms of demographic and school-related dimensions are summarized in table~\ref{tab:fraq_stats} in \ref{sec:appendIndCharac}. 

\subsubsection{Measurement toolkit}
\label{sec:measureTool}

The measurement toolkit included two main parts of measurement. The former referred to a profile assessment and the latter to a Kidlearn intervention assessment. A timeline was designed to articulate these assessments around the Kidlearn game inside each session (see Table~\ref{table:exp3orga}). 

\subsubsubsection{Profile metrics}
\label{sec:profileMeasure}

To have an assessment of the background of each participant, several measures have been collected. 
First, the General Profile (GP) questionnaire (\ref{ques:GP}) refers to questions related to student information such as gender, experience with technologies, his perception and habits to use money (simple manipulation or money calculation). Another set of questions concerns the habits of choosing in life-related decision making such as clothes or food choices (self-choice score): it is used to probe the student's self-determination trait\footnote[1]{When a behaviour is self-determined, the regulatory process is choice, but when it is controlled, the regulatory process is compliance (or in some cases defiance) \cite{SRdeci1991motivation}.}.

Second, a school-related psychological perception assessment has been carried from the use of two questionnaires,i.e., the Quality of School-Life Scale  (QSLS)\cite{SRlazar1999quality} and the Learner Empowerment Scale (LES)\cite{SRweber2005student}. 

The QSLS evaluates the quality of school life experienced by the student (satisfaction at school, the student's interest in academic learning, and the nature of the student-teacher interactions / students' attitude towards the teacher), and it's a high predictor of disengagement behaviours in school. 

The LES measures the learner's empowerment with questions such as "This course will help me achieve my future goals" and "I have the qualifications to succeed in this class". The two questionnaires were combined and reworked to produce the School Profile (SP) questionnaire (\ref{ques:SP}), which contains 10 items on a 5-point Likert scale ; 5 items are related to school while the others to relationships.  

The overall profile metrics aimed to establish an initial profile for each student in order to have equivalent control and experimental groups in respect of personal factors that may affect the results of our experiment such as demographic factors, technology experience, everyday self-determined behaviors, money manipulation and calculation and school experience and perceptions (see table~\ref{tab:fraq_stats}). 

\subsubsubsection{Kidlearn Intervention Assessment}

To assess Kidlearn Intervention according to the 4 experimental conditions, the assessment was entailed three parts. The first one corresponded to learning activity data from direct interactions with the KidLearn game computed as an "Activity score". The second part is dedicated to the learning effectiveness assessment of the Kidlearn intervention with pre-and post-test regarding calculation performance before and after the Kidlearn intervention. Finally, the third part included assessments regarding the learning experience elicited across the sessions of Kidlearn intervention (emotional and motivation scales for probing the learning experience according to the four manipulated conditions)

\subsubsubsection{Activity Score from KidLearn Game}
\label{sec:IngameMeas}

The exploitation of the data from the interaction of the students with the Kidlearn activity is done using two kinds of indicators tracing the learning path of each student.  
The first one (Fig.~\ref{fig:activityScores}) is used to compare quantitatively the activities made by the students. Here, two scores are built, and used in result section. The former represents the activities reached by a student in the activity space, and the latter one represents the success rate over these reached activities. For a student, these scores are defined as follows:
\begin{equation}
  score^{reached}(t) = \sum^{4}_{i=1} max(\{l^{i,j}(t)~|~j \in L^i \}) f^i   
\label{eq:scoreReachExp3}
\end{equation}
\begin{equation}
  score^{success}(t) = \sum^{4}_{i=1} max(\{\delta_{i,j}(t) l^{i,j}(t)~|~j \in L^i\}) f^i  
\label{eq:scoreSuccExp3}
\end{equation}

\noindent where $i$ is the index corresponding to each type of activity, $f^i$ is the factor related to the activity type $i$ as described in table \ref{tab:scoreTable}. If the level $j$ for type $i$ has been reached at time $t$, then $l^{i,j}(t) = j$, or else $l^{i,j}(t) = 0$. $\delta_{i,j}(t)$ is the student's success rate over the $4$ last steps activity type $i$, level $j$. For example, at time $t$, if a student has reached M$4$, MM$2$, R$1$ and did not reach RM, his $score^{reached}$ is equal to : $4 \times 1 + 2 \times 2 + 1 \times 3 + 0 \times 4 = 11$.

\begin{table}[h]
\centering
  \begin{tabular}{c|c|c|c|c}
    \hline
    Type  & M  & MM & R & RM    \\\hline
    Index $i$   & 1  & 2 & 3 & 4    \\ 
    Factor $f^i$  & 1  & 2 & 3 & 4    \\ 
    Levels $L^i$  & 1-6  & 0-4 & 0-4 & 0-4    \\\hline
 
  \end{tabular}
  
  \caption[Score Factor]{Table of factor, index and the number of levels for each type of activity to compute scores in equations~\ref{eq:scoreReachExp3} and~\ref{eq:scoreSuccExp3}. The level $0$ represents the fact a student has not made any exercise of this type yet. Students start with exercises of type M, so there is no level 0}
  \label{tab:scoreTable}
\end{table}

The second indicator (Fig.~\ref{fig:magHistoExp3TimeAct}) dynamically traced the curriculum of each student in terms of activities performed during Kidlearn sessions. Precisely, for a given time step $t$ and condition, a matrix slot represents the state of an activity (ordinate) for a particular student (abscissa). A slot is coded in grey if a student has never explored the corresponding activity and it is coded in purple if the student is doing this activity at time $t$. When a student has explored an activity, the slot is coded of tint of green depending on the student's success rate (light green: low, dark green: high).

\subsubsubsection{KidLearn Learning Score (pre-post effect)}
\label{sec:LearningMeas}

A math-test, used as a pre-post metric to measure the students money-related computational learning, has been made by a pedagogical expert. The pre- and post-test are composed of 20 items scoring from 0 to 1 (max score is 20). The three first questions are general mathematical questions about composition of numbers. The 17 other questions are related to the manipulation money, composition of numbers, addition and subtraction, and are directly related to the activities of the Kidlearn scenario. The pre-test happens at the beginning of the first session, while the post-test happens at the end of the last session. 

The pre- and post-test are presented on the tablet on a dedicated interface (different from the ITS one). Each item of the test evaluates the student over knowledge and skills related to money manipulation, number composition, addition or subtraction (similar to the skills and knowledge trained in the ITS). Both tests include the same items organised in the same order but the items' wording have randomly selected values for each item and each student (with verification that no items in the post-test have the same values in their wording as the ones in the pre-test for one student).

This way, each question is generated with a unique structure but with random values set to have the same mathematical difficulty with the goal to reduce learning effect (test-retest effect) on learning score due to the test repetition (pre-post test). This also allowed  to reduce the possibilities of student cheating. 

\subsubsubsection{Kidlearn-related Learning experience}
\label{sec:motivMeasure}

\begin{figure}
\centering
  \includegraphics[width=0.7\linewidth]{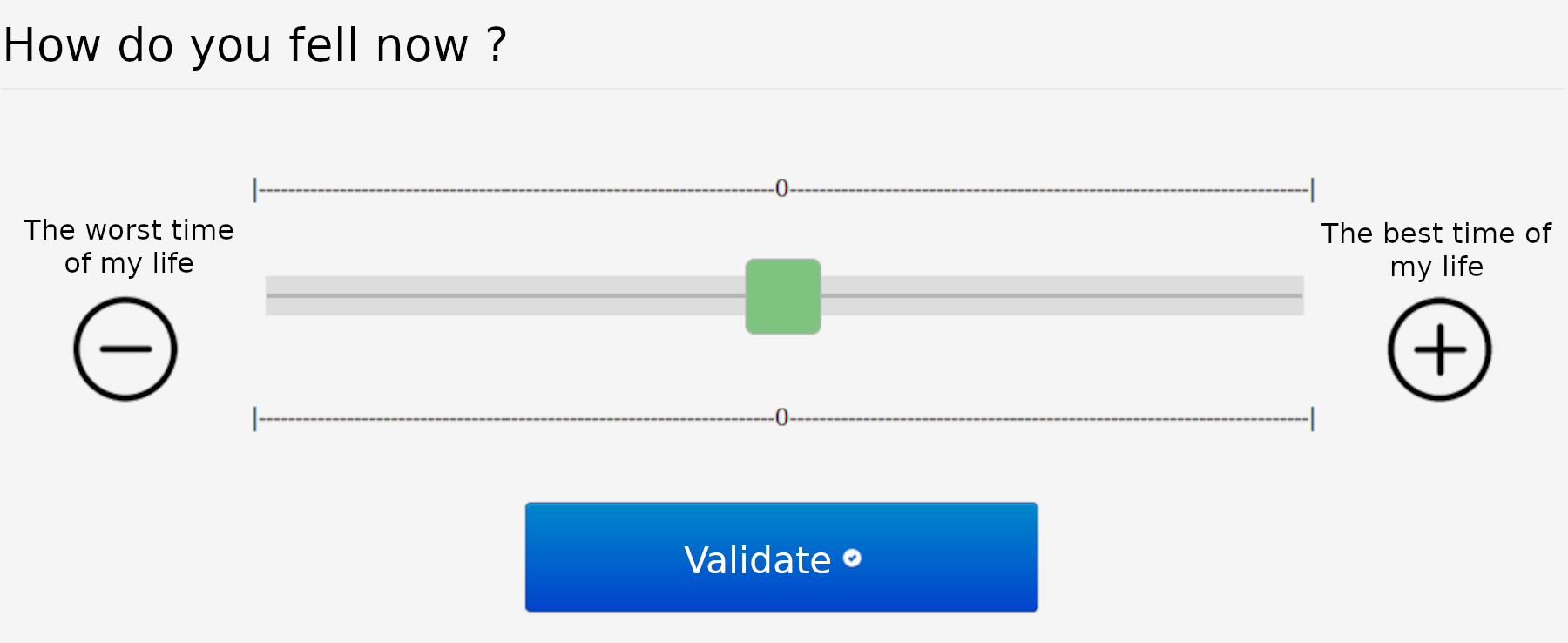}
  \caption[well-being Scale]{\label{fig:weelBeingScale} Emotional scale. The student is asked a question ``How do you feel ?'' and moves towards + (best time) or - (worst time) depending on how he/she feels right now.}
\end{figure}

Two questionnaires rates learning experience from Kidlearn game, using an emotional scale and a motivation questionnaire. 
The first one assessed the emotional experience or well-being related to Kidlearn practice. It consisted of self-measurements of emotional valence elicited during each Kidlearn session. 
This self-measurement is a simplified version of the Self-Assessment Manikin \cite{SRbradley1994measuring}. It is presented in figure~\ref{fig:weelBeingScale}. 

The student must position a cursor between "the best time of my life" (intensive and positive emotional state) and "the worst time of my life" (intensive and negative emotional state) depending on how he/she feels right now (it is scored from -50 to 50). They have to answer to this scale at the beginning, middle and end of each one of the four sessions. A summative score is computed from all inputs (used in result section),
max score is 600 across the four sessions. This provides us an approximate measure of the evolution of the well-being of the children during each session. 

The second measure is a motivation questionnaire from Vallerand's questionnaire \cite{SRvallerand1992academic, SRvallerand1989construction} providing a measure of the amount of student motivation. It is based on Self-Determination Theory\cite{SRdeci1985general} and is commonly used to ascertain the elicitation of intrinsic and extrinsic motivation\cite{SRdesrochers2006elaboration}. Once adapted to a public of children users playing a serious game, children had to answer a 21 items-questionnaire on their experience playing during the last session (max score of global motivation is 21). This questionnaire, entitled Type of Motivation (TM), is presented in \ref{ques:TM}. 

All this measurement toolkit was integrated into the interface of the web application used in the experiment. Hence, no paper-pencil assessment was used, and then the data collecting was fully computerized to standardize the collecting procedure and to simplify after both the data base structuring and its statistical analysis . 

\subsubsection{Study Procedure}  
\label{sec:toolDesign}

The  study procedure followed several steps managed by two experimenters. Precisely, four experimental sessions are organized over two weeks where the three Kidlearn game phases (30 mn/phase), the profile metrics as well as Kidlearn related assessment were performed. Each participant had to undergo the four successive sessions according to a specific organisation of its contents. 
The organisation of contents of each session is presented in table~\ref{table:exp3orga}.
\begin{table*}[ht!]
    \centering
    \begin{tabular}{l|l|l|l}
   Session 1 ($\sim$1h20)     & Session 2 ($\sim$40 min)    & Session 3 ($\sim$40 min)    & Session 4 ($\sim$40min) \\\hline
     1. Project explanation & 1. Emotional scale   & 1. Emotional scale   & 1. Emotional scale \\
     2. Emotional scale  & 2. Game phase (30 min)  & 2. SP questionnaire     & 2. Motivation questionnaire \\
     3. GP questionnaire  & 3. Emotional scale   & 3. Emotional scale   & 3. Emotional scale  \\
     4. Math pre-test (20 min) & 4. GI questionnaire & 4. Game phase (30 min)  & 4. Math post-test (20 min) \\  
     5. Emotional scale  & 5. Emotional scale  & 5. Emotional scale 3   & 5. Emotional scale \\
     6. Game phase (30 min) &  &    &  \\
     7. Emotional scale & & &\\\hline

    \end{tabular}
  
  \caption[Sessions Timeline]{Sessions Timeline. The table shows the sequence of steps for each session. There are 7 steps in the first session and 5 steps in the other sessions. The GI questionnaire is a hand made questionnaire done to evaluate our Kidlearn interface, results of this questionnaire is not include in the study due to technical problem in collecting data.}
  \label{table:exp3orga}

\end{table*}

There are some points to consider in particular. 
A game phase is always preceded by a self-emotional assessment and followed by a self-emotional assessment scale. 
The questionnaires were distributed over the 4 sessions so as not to ask too many questions at a time. 
In the first session, the General Profile (GP) questionnaire is done first. They have all the time they want to answer. It allows students to acclimatize to the tablet, the web site interface and to be confident in the tablet usage. After, the Math pre-test followed by a game phase were administrated. The session 2 and 3 are similarly structured except the exclusion of contents related to project explanation and GP profile. Finally, the session 4 is dedicated to post intervention assessment including the motivation questionnaire and the Math post-test.

For providing an optimal learning environment for the experiment, each class group was divided into two sub-groups for the whole of sessions.Each sub-group was doing the same session in parallel in different rooms supervised by a researcher. 
At each step of the session, finishing a step leads the student to the waiting page. While waiting for the others to finish the step, students can draw on their draft or if waiting time may be long (as for tests, some finished what they can do in 10 minutes), they can read a book present in the classroom or discuss with other that also finished without disturbing the classroom. This process allowed to considerably reduce distraction and provided a better and quieter working environment for the students.
\subsection{Population characteristics for each condition}
\label{sec:appendIndCharac}
Table \ref{tab:fraq_stats} shows the student's frequencies for gender and calculation liking as well as the mean and standard deviation for each group (Predef, PCO, ZPDES and ZCO condition) regarding the studied profile metrics (Self-choice, Technology experience, Money manipulation, Money calculation and School-related perception). Group comparisons did not reach the significance ($p > .05$). 

\begin{table*}[ht!]
    \centering
    \begin{tabular}{lcc||cc||cc||cc}
                        & \multicolumn{2}{c}{Predef} & \multicolumn{2}{c}{PCO} & \multicolumn{2}{c}{ZPDES} & \multicolumn{2}{c}{ZCO} \\\hline
        \multirow{2}{*}{Gender}          & F & M & F & M & F & M & F & M  \\
                            & 31 & 31 & 29 & 30 & 38 & 38 & 46 & 22   \\\hline
        \multirow{2}{*}{Like calculation}          & Yes & No & Yes & No& Yes & No& Yes & No  \\
                            & 53 & 9 & 54 & 5 & 68 & 8 & 53 & 15   \\\hline
                            & Mean  & SD   & Mean & SD & Mean & SD & Mean & SD  \\
        Self-choice score   & 9.39  & 3.19 & 8.73 & 2.97 & 8.82 & 2.75 & 9.10 & 2.53    \\\hline
        Technology Experience   & 3.05  & 1.84 & 3.05 & 1.83 & 3.51 & 1.69 & 3.19 & 1.73    \\\hline
        Money Manipulation  & 2.45  & 1.00 & 2.37 & 0.89 & 2.38 & 0.89 & 2.50 & 0.82    \\\hline
        Money Calculation   & 4.16  & 1.23 & 4.16 & 1.08 & 4.17 & 1.11 & 4.06 & 1.26    \\\hline
        School Profile      & 27.32 & 6.15 & 27.32& 6.79 & 28.53& 6.11 & 28.62& 5.53    \\\hline

    \end{tabular}
    \caption{Population characteristics for each condition}
    \label{tab:fraq_stats}
\end{table*}
%

%% file: Appendix.tex
\section{Appendix}
\label{sec:appendix}


\subsection{Questionnaires}
\label{sec:ques}

\subsubsection{Type of Motivation Questionnaire (TM)}
\label{ques:TM}
The children can answer by yes or no to the items.
\begin{enumerate}
    \item I played the game because I would have felt ashamed if I had not tried.
    \item I played the game because I like succeeding at a game.
    \item I played the game, but I do not know why.
    \item I played the game because it makes me happy when I correctly answer a challenging activity.
    \item I played the game to avoid doing an exercise the teacher would have given me.
    \item I played the game because I am happy to learn how to handle money.
    \item I played the game because it relaxes me when I play.
    \item I played the game to be able to use the tablet/computer.
    \item I played the game because I did what I was told to do.
    \item I played the game because I am happy when I play.
    \item I played the game to show that I can do things.
    \item I played the game because I think that if I can handle money better, I can help a friend or my parents with it.
    \item I played the game because I never get bored when I play.
    \item I played the game because I like learning new things.
    \item I played the game because it is a way for me to improve my calculations.
    \item I do not know why I played the game; I feel like I was bored.
    \item I played the game because I learn many things that interest me.
    \item I played the game because I have always done well in games and I want to prove it.
    \item I played the game to get a good score.
    \item I played the game because it puts me in a good mood when I succeed.
    \item I played the game because I am happy when I feel like I am getting better at a game.
\end{enumerate}

\subsubsection{General Profile (GP)}
\label{ques:GP}
\begin{enumerate}
    \item About gender:
    \begin{itemize}
        \item (a) I am a boy
        \item (b) I am a girl
    \end{itemize}
    \item About the use of money:
    \begin{itemize}
        \item (a) I have used money and I like it
        \item (b) I have used money but I do not like it
        \item (c) I have never used money but I would like to learn
        \item (d) I have never used money and I do not want to learn
    \end{itemize}
    \item Do you like doing calculations?
    \begin{itemize}
        \item (a) No
        \item (b) Yes
    \end{itemize}
    \item What screens do you use at home?
    \begin{itemize}
        \item (a) A computer
        \item (b) A tablet
        \item (c) A smartphone
        \item (d) A portable console
        \item (e) A home console
    \end{itemize}
    \item How often do you use these screens?
    \begin{itemize}
        \item (a) Never
        \item (b) Only on weekends
        \item (c) A few times a week
        \item (d) Every day
    \end{itemize}
    \item At home, I choose the activities I do:
    \begin{itemize}
        \item (a) Not at all
        \item (b) Not often
        \item (c) Sometimes yes, sometimes no
        \item (d) Often
        \item (e) All the time
    \end{itemize}
    \item Generally, I choose the food I eat:
    \begin{itemize}
        \item (a) Not at all
        \item (b) Not often
        \item (c) Sometimes yes, sometimes no
        \item (d) Often
        \item (e) All the time
    \end{itemize}
    \item I choose who my friends are:
    \begin{itemize}
        \item (a) Not at all
        \item (b) Not often
        \item (c) Sometimes yes, sometimes no
        \item (d) Often
        \item (e) All the time
    \end{itemize}
    \item I choose the clothes I wear:
    \begin{itemize}
        \item (a) Not at all
        \item (b) Not often
        \item (c) Sometimes yes, sometimes no
        \item (d) Often
        \item (e) All the time
    \end{itemize}
\end{enumerate}

\subsubsection{School Profile Questionnaire (SP)}
\label{ques:SP}
\begin{enumerate}
    \item I do interesting things in class:
    \begin{itemize}
        \item (a) Never
        \item (b) Almost never
        \item (c) Sometimes
        \item (d) Often
        \item (e) All the time
    \end{itemize}
    \item At school, I get along with:
    \begin{itemize}
        \item (a) Nobody
        \item (b) One or two people
        \item (c) Less than half
        \item (d) More than half
        \item (e) Everyone
    \end{itemize}
    \item I like my teacher:
    \begin{itemize}
        \item (a) Not at all
        \item (b) Not much
        \item (c) A little
        \item (d) A lot
        \item (e) Really a lot
    \end{itemize}
    \item At school, I feel bored:
    \begin{itemize}
        \item (a) All the time
        \item (b) Often
        \item (c) Sometimes
        \item (d) Almost never
        \item (e) Never
    \end{itemize}
    \item If school were a person, it would be:
    \begin{itemize}
        \item (a) My enemy
        \item (b) A stranger
        \item (c) Someone I know
        \item (d) A friend
        \item (e) My best friend
    \end{itemize}
\end{enumerate}